\documentclass[twoside,11pt,a4paper]{article}
\usepackage{color,pstcol}
\begin{document}

%
%
%



\newcommand{\ysssrcnb}
          {$\Ysssrcnb$}
\newcommand{\Ysssrcnb}
           {N}

\newcommand{\ysssensnb}
          {$\Ysssensnb$}
\newcommand{\Ysssensnb}
           {P}

\newcommand{\yssoutsepsystsigassocconttimecentnb}
{$\Yssoutsepsystsigassocconttimecentnb$}
\newcommand{\Yssoutsepsystsigassocconttimecentnb}
{L}

\newcommand{\yssbothtimevalnb}
          {$\Yssbothtimevalnb$}
\newcommand{\Yssbothtimevalnb}
           {L}

\newcommand{\yssbothfreqvalnb}
          {$\Yssbothfreqvalnb$}
\newcommand{\Yssbothfreqvalnb}
           {\Yssbothtimevalnb^{\prime}}

\newcommand{\yssbothscalevalnb}
          {$\Yssbothscalevalnb$}
\newcommand{\Yssbothscalevalnb}
           {\Yssbothtimevalnb^{\prime \prime}}



\newcommand{\yssconttimeval}
          {$\Yssconttimeval$}
\newcommand{\Yssconttimeval}
           {t}

\newcommand{\yssconttimevalsubone}
          {$\Yssconttimevalsubone$}
\newcommand{\Yssconttimevalsubone}
           {\Yssconttimeval _1}

\newcommand{\yssconttimevalsubtwo}
          {$\Yssconttimevalsubtwo$}
\newcommand{\Yssconttimevalsubtwo}
           {\Yssconttimeval _2}

\newcommand{\ysssubconttimeval}
          {$\Ysssubconttimeval$}
\newcommand{\Ysssubconttimeval}
           {p}

\newcommand{\yssconttimevalsubstd}
          {$\Yssconttimevalsubstd$}
\newcommand{\Yssconttimevalsubstd}
           {\Yssconttimeval _{\Ysssubconttimeval}}

\newcommand{\ysssubotherconttimeval}
          {$\Ysssubotherconttimeval$}
\newcommand{\Ysssubotherconttimeval}
           {q}

\newcommand{\yssconttimevalsubstdother}
          {$\Yssconttimevalsubstdother$}
\newcommand{\Yssconttimevalsubstdother}
           {\Yssconttimeval _{\Ysssubotherconttimeval}}

\newcommand{\yssconttimevalother}
          {$\Yssconttimevalother$}
\newcommand{\Yssconttimevalother}
           {t^{\prime}}

\newcommand{\yssconttimevalothersubone}
          {$\Yssconttimevalothersubone$}
\newcommand{\Yssconttimevalothersubone}
           {\Yssconttimevalother _1}

\newcommand{\yssconttimevalothersubtwo}
          {$\Yssconttimevalothersubtwo$}
\newcommand{\Yssconttimevalothersubtwo}
           {\Yssconttimevalother _2}





\newcommand{\yssdisctimeval}
          {$\Yssdisctimeval$}
\newcommand{\Yssdisctimeval}
           {n}

\newcommand{\yssdisctimevalsubone}
          {$\Yssdisctimevalsubone$}
\newcommand{\Yssdisctimevalsubone}
           {\Yssdisctimeval _1}

\newcommand{\yssdisctimevalsubtwo}
          {$\Yssdisctimevalsubtwo$}
\newcommand{\Yssdisctimevalsubtwo}
           {\Yssdisctimeval _2}

\newcommand{\ysssubdisctimeval}
          {$\Ysssubdisctimeval$}
\newcommand{\Ysssubdisctimeval}
           {i}

\newcommand{\yssdisctimevalsubstd}
          {$\Yssdisctimevalsubstd$}
\newcommand{\Yssdisctimevalsubstd}
           {\Yssdisctimeval _{\Ysssubdisctimeval}}

\newcommand{\ysssubotherdisctimeval}
          {$\Ysssubotherdisctimeval$}
\newcommand{\Ysssubotherdisctimeval}
           {j}

\newcommand{\yssdisctimevalsubstdother}
          {$\Yssdisctimevalsubstdother$}
\newcommand{\Yssdisctimevalsubstdother}
           {\Yssdisctimeval _{\Ysssubotherdisctimeval}}

\newcommand{\yssdisctimevalother}
          {$\Yssdisctimevalother$}
\newcommand{\Yssdisctimevalother}
           {m}

\newcommand{\yssdisctimevalothersubone}
          {$\Yssdisctimevalothersubone$}
\newcommand{\Yssdisctimevalothersubone}
           {\Yssdisctimevalother _1}

\newcommand{\yssdisctimevalothersubtwo}
          {$\Yssdisctimevalothersubtwo$}
\newcommand{\Yssdisctimevalothersubtwo}
           {\Yssdisctimevalother _2}


\newcommand{\yssdisctimelag}
          {$\Yssdisctimelag$}
\newcommand{\Yssdisctimelag}
           {l}

\newcommand{\yssdisctimelagsubone}
          {$\Yssdisctimelagsubone$}
\newcommand{\Yssdisctimelagsubone}
           {\Yssdisctimelag _1}

\newcommand{\yssdisctimelagsubtwo}
          {$\Yssdisctimelagsubtwo$}
\newcommand{\Yssdisctimelagsubtwo}
           {\Yssdisctimelag _2}


\newcommand{\ysscontfreqval}
          {$\Ysscontfreqval$}
\newcommand{\Ysscontfreqval}
           {\omega}

\newcommand{\ysssubcontfreqval}
          {$\Ysssubcontfreqval$}
\newcommand{\Ysssubcontfreqval}
           {\Ysssubconttimeval^{\prime}}

\newcommand{\ysscontfreqvalsubstd}
          {$\Ysscontfreqvalsubstd$}
\newcommand{\Ysscontfreqvalsubstd}
           {\Ysscontfreqval _{\Ysssubcontfreqval}}



\newcommand{\ysstimescalecontshiftval}
          {$\Ysstimescalecontshiftval$}
\newcommand{\Ysstimescalecontshiftval}
           {%
       \tau
       }

\newcommand{\ysstimescalecontscaleval}
          {$\Ysstimescalecontscaleval$}
\newcommand{\Ysstimescalecontscaleval}
           {%
       d
       }

\newcommand{\ysstimescalecontscaleexp}
          {$\Ysstimescalecontscaleexp$}
\newcommand{\Ysstimescalecontscaleexp}
           {%
       j
       }

\newcommand{\ysssubtimescalecontscaleval}
          {$\Ysssubtimescalecontscaleval$}
\newcommand{\Ysssubtimescalecontscaleval}
           {\Ysssubconttimeval^{\prime \prime}}

\newcommand{\ysstimescalecontscalevalsubstd}
          {$\Ysstimescalecontscalevalsubstd$}
\newcommand{\Ysstimescalecontscalevalsubstd}
           {\Ysstimescalecontscaleval _{\Ysssubtimescalecontscaleval}}

\newcommand{\ysstimescalecontmotherwav}
          {$\Ysstimescalecontmotherwav$}
\newcommand{\Ysstimescalecontmotherwav}
           {\psi}

\newcommand{\ysstimescalecontmotherwavval}
          {$\Ysstimescalecontmotherwavval$}
\newcommand{\Ysstimescalecontmotherwavval}
           {\Ysstimescalecontmotherwav ( \Yssconttimeval )}

\newcommand{\ysstimescalecontscaleshiftwav}
          {$\Ysstimescalecontscaleshiftwav$}
\newcommand{\Ysstimescalecontscaleshiftwav}
           {\Ysstimescalecontmotherwav
        _{\Ysstimescalecontshiftval , \Ysstimescalecontscaleval}
       }

\newcommand{\ysstimescalecontscaleshiftwavval}
          {$\Ysstimescalecontscaleshiftwavval$}
\newcommand{\Ysstimescalecontscaleshiftwavval}
           {\Ysstimescalecontscaleshiftwav ( \Yssconttimeval )}

\newcommand{\ysstimescalecontcoefnot}
          {$\Ysstimescalecontcoefnot$}
\newcommand{\Ysstimescalecontcoefnot}
           {W}


\newcommand{\ysssepsystlag}
          {$\Ysssepsystlag$}
\newcommand{\Ysssepsystlag}
           {k}

\newcommand{\ysscomplfiltlag}
          {$\Ysscomplfiltlag$}
\newcommand{\Ysscomplfiltlag}
           {m}

\newcommand{\ysscomplfiltlagother}
          {$\Ysscomplfiltlagother$}
\newcommand{\Ysscomplfiltlagother}
           {l}


\newcommand{\ysswayindex}
          {$\Ysswayindex$}
\newcommand{\Ysswayindex}
           {i}

\newcommand{\ysswayindexother}
          {$\Ysswayindexother$}
\newcommand{\Ysswayindexother}
           {j}

\newcommand{\ysswayindexthird}
          {$\Ysswayindexthird$}
\newcommand{\Ysswayindexthird}
           {k}

\newcommand{\ysswayindexfourth}
          {$\Ysswayindexfourth$}
\newcommand{\Ysswayindexfourth}
           {l}




\newcommand{\ysssrcsiginnovdisctimecentvec}
          {$\Ysssrcsiginnovdisctimecentvec$}
\newcommand{\Ysssrcsiginnovdisctimecentvec}
           {p}

\newcommand{\ysssrcsiginnovdisctimecentvecval}
          {$\Ysssrcsiginnovdisctimecentvecval$}
\newcommand{\Ysssrcsiginnovdisctimecentvecval}
           {\Ysssrcsiginnovdisctimecentvec (\Yssdisctimeval)}

\newcommand{\ysssrcsiginnnovdisctimecentone}
          {$\Ysssrcsiginnovdisctimecentone$}
\newcommand{\Ysssrcsiginnovdisctimecentone}
           {\Ysssrcsiginnovdisctimecentvec _{1}}

\newcommand{\ysssrcsiginnovdisctimecentoneval}
          {$\Ysssrcsiginnovdisctimecentoneval$}
\newcommand{\Ysssrcsiginnovdisctimecentoneval}
           {\Ysssrcsiginnovdisctimecentone (\Yssdisctimeval)}

\newcommand{\ysssrcsiginnnovdisctimecenttwo}
          {$\Ysssrcsiginnovdisctimecenttwo$}
\newcommand{\Ysssrcsiginnovdisctimecenttwo}
           {\Ysssrcsiginnovdisctimecentvec _{2}}

\newcommand{\ysssrcsiginnovdisctimecenttwoval}
          {$\Ysssrcsiginnovdisctimecenttwoval$}
\newcommand{\Ysssrcsiginnovdisctimecenttwoval}
           {\Ysssrcsiginnovdisctimecenttwo (\Yssdisctimeval)}

\newcommand{\ysssrcsiginnovdisctimecentindex}
          {$\Ysssrcsiginnovdisctimecentindex$}
\newcommand{\Ysssrcsiginnovdisctimecentindex}
           {\Ysssrcsiginnovdisctimecentvec _{\Ysswayindex}}

\newcommand{\ysssrcsiginnovdisctimecentindexval}
          {$\Ysssrcsiginnovdisctimecentindexval$}
\newcommand{\Ysssrcsiginnovdisctimecentindexval}
           {\Ysssrcsiginnovdisctimecentindex (\Yssdisctimeval)}

\newcommand{\ysssrcsiginnovdisctimecentindexother}
          {$\Ysssrcsiginnovdisctimecentindexother$}
\newcommand{\Ysssrcsiginnovdisctimecentindexother}
           {\Ysssrcsiginnovdisctimecentvec _{\Ysswayindexother}}

\newcommand{\ysssrcsiginnovdisctimecentindexotherval}
          {$\Ysssrcsiginnovdisctimecentindexotherval$}
\newcommand{\Ysssrcsiginnovdisctimecentindexotherval}
           {\Ysssrcsiginnovdisctimecentindexother (\Yssdisctimeval)}


\newcommand{\ysssrcsiginnovdisctimecentveczt}
          {$\Ysssrcsiginnovdisctimecentveczt$}
\newcommand{\Ysssrcsiginnovdisctimecentveczt}
           {P}

\newcommand{\ysssrcsiginnovdisctimecentvecztval}
          {$\Ysssrcsiginnovdisctimecentvecztval$}
\newcommand{\Ysssrcsiginnovdisctimecentvecztval}
           {\Ysssrcsiginnovdisctimecentveczt (z)}

\newcommand{\ysssrcsiginnovdisctimecentindexzt}
          {$\Ysssrcsiginnovdisctimecentindexzt$}
\newcommand{\Ysssrcsiginnovdisctimecentindexzt}
           {\Ysssrcsiginnovdisctimecentveczt _{\Ysswayindex}}

\newcommand{\ysssrcsiginnovdisctimecentindexztval}
          {$\Ysssrcsiginnovdisctimecentindexztval$}
\newcommand{\Ysssrcsiginnovdisctimecentindexztval}
           {\Ysssrcsiginnovdisctimecentindexzt (z)}



\newcommand{\ysssrcsigconttimecentvec}
          {$\Ysssrcsigconttimecentvec$}
\newcommand{\Ysssrcsigconttimecentvec}
           {s}

\newcommand{\ysssrcsigconttimecentvecval}
          {$\Ysssrcsigconttimecentvecval$}
\newcommand{\Ysssrcsigconttimecentvecval}
           {\Ysssrcsigconttimecentvec (\Yssconttimeval)}

\newcommand{\ysssrcsigconttimecentone}
          {$\Ysssrcsigconttimecentone$}
\newcommand{\Ysssrcsigconttimecentone}
           {\Ysssrcsigconttimecentvec _{1}}

\newcommand{\ysssrcsigconttimecentoneval}
          {$\Ysssrcsigconttimecentoneval$}
\newcommand{\Ysssrcsigconttimecentoneval}
           {\Ysssrcsigconttimecentone (\Yssconttimeval)}

\newcommand{\ysssrcsigconttimecenttwo}
          {$\Ysssrcsigconttimecenttwo$}
\newcommand{\Ysssrcsigconttimecenttwo}
           {\Ysssrcsigconttimecentvec _{2}}

\newcommand{\ysssrcsigconttimecenttwoval}
          {$\Ysssrcsigconttimecenttwoval$}
\newcommand{\Ysssrcsigconttimecenttwoval}
           {\Ysssrcsigconttimecenttwo (\Yssconttimeval)}

\newcommand{\ysssrcsigconttimecentthree}
          {$\Ysssrcsigconttimecentthree$}
\newcommand{\Ysssrcsigconttimecentthree}
           {\Ysssrcsigconttimecentvec _{3}}

\newcommand{\ysssrcsigconttimecentthreeval}
          {$\Ysssrcsigconttimecentthreeval$}
\newcommand{\Ysssrcsigconttimecentthreeval}
           {\Ysssrcsigconttimecentthree (\Yssconttimeval)}

\newcommand{\ysssrcsigconttimecentindex}
          {$\Ysssrcsigconttimecentindex$}
\newcommand{\Ysssrcsigconttimecentindex}
           {\Ysssrcsigconttimecentvec _{\Ysswayindex}}

\newcommand{\ysssrcsigconttimecentindexval}
          {$\Ysssrcsigconttimecentindexval$}
\newcommand{\Ysssrcsigconttimecentindexval}
           {\Ysssrcsigconttimecentindex (\Yssconttimeval)}

\newcommand{\ysssrcsigconttimecentindexother}
          {$\Ysssrcsigconttimecentindexother$}
\newcommand{\Ysssrcsigconttimecentindexother}
           {\Ysssrcsigconttimecentvec _{\Ysswayindexother}}

\newcommand{\ysssrcsigconttimecentindexotherval}
          {$\Ysssrcsigconttimecentindexotherval$}
\newcommand{\Ysssrcsigconttimecentindexotherval}
           {\Ysssrcsigconttimecentindexother (\Yssconttimeval)}

\newcommand{\ysssrcsigconttimecentindexthird}
          {$\Ysssrcsigconttimecentindexthird$}
\newcommand{\Ysssrcsigconttimecentindexthird}
           {\Ysssrcsigconttimecentvec _{\Ysswayindexthird}}

\newcommand{\ysssrcsigconttimecentindexthirdval}
          {$\Ysssrcsigconttimecentindexthirdval$}
\newcommand{\Ysssrcsigconttimecentindexthirdval}
           {\Ysssrcsigconttimecentindexthird (\Yssconttimeval)}

\newcommand{\ysssrcsigconttimecentindexfourth}
          {$\Ysssrcsigconttimecentindexfourth$}
\newcommand{\Ysssrcsigconttimecentindexfourth}
           {\Ysssrcsigconttimecentvec _{\Ysswayindexfourth}}

\newcommand{\ysssrcsigconttimecentindexfourthval}
          {$\Ysssrcsigconttimecentindexfourthval$}
\newcommand{\Ysssrcsigconttimecentindexfourthval}
           {\Ysssrcsigconttimecentindexfourth (\Yssconttimeval)}

\newcommand{\ysssrcsigconttimecentlast}
          {$\Ysssrcsigconttimecentlast$}
\newcommand{\Ysssrcsigconttimecentlast}
           {\Ysssrcsigconttimecentvec _{\Ysssrcnb}}

\newcommand{\ysssrcsigconttimecentlastval}
          {$\Ysssrcsigconttimecentlastval$}
\newcommand{\Ysssrcsigconttimecentlastval}
           {\Ysssrcsigconttimecentlast (\Yssconttimeval)}


\newcommand{\ysssrcsigassocconttimecentvec}
          {$\Ysssrcsigassocconttimecentvec$}
\newcommand{\Ysssrcsigassocconttimecentvec}
           {\Ysssrcsigconttimecentvec ^{\prime}}

\newcommand{\ysssrcsigassocconttimecentvecval}
          {$\Ysssrcsigassocconttimecentvecval$}
\newcommand{\Ysssrcsigassocconttimecentvecval}
           {\Ysssrcsigassocconttimecentvec (\Yssconttimeval)}

\newcommand{\ysssrcsigassocconttimecentone}
          {$\Ysssrcsigassocconttimecentone$}
\newcommand{\Ysssrcsigassocconttimecentone}
           {\Ysssrcsigassocconttimecentvec _{1}}

\newcommand{\ysssrcsigassocconttimecentoneval}
          {$\Ysssrcsigassocconttimecentoneval$}
\newcommand{\Ysssrcsigassocconttimecentoneval}
           {\Ysssrcsigassocconttimecentone (\Yssconttimeval)}

\newcommand{\ysssrcsigassocconttimecenttwo}
          {$\Ysssrcsigassocconttimecenttwo$}
\newcommand{\Ysssrcsigassocconttimecenttwo}
           {\Ysssrcsigassocconttimecentvec _{2}}

\newcommand{\ysssrcsigassocconttimecenttwoval}
          {$\Ysssrcsigassocconttimecenttwoval$}
\newcommand{\Ysssrcsigassocconttimecenttwoval}
           {\Ysssrcsigassocconttimecenttwo (\Yssconttimeval)}

\newcommand{\ysssrcsigassocconttimecentthree}
          {$\Ysssrcsigassocconttimecentthree$}
\newcommand{\Ysssrcsigassocconttimecentthree}
           {\Ysssrcsigassocconttimecentvec _{3}}

\newcommand{\ysssrcsigassocconttimecentthreeval}
          {$\Ysssrcsigassocconttimecentthreeval$}
\newcommand{\Ysssrcsigassocconttimecentthreeval}
           {\Ysssrcsigassocconttimecentthree (\Yssconttimeval)}

\newcommand{\ysssrcsigassocconttimecentindex}
          {$\Ysssrcsigassocconttimecentindex$}
\newcommand{\Ysssrcsigassocconttimecentindex}
           {\Ysssrcsigassocconttimecentvec _{\Ysswayindex}}

\newcommand{\ysssrcsigassocconttimecentindexval}
          {$\Ysssrcsigassocconttimecentindexval$}
\newcommand{\Ysssrcsigassocconttimecentindexval}
           {\Ysssrcsigassocconttimecentindex (\Yssconttimeval)}

\newcommand{\ysssrcsigassocconttimecentindexother}
          {$\Ysssrcsigassocconttimecentindexother$}
\newcommand{\Ysssrcsigassocconttimecentindexother}
           {\Ysssrcsigassocconttimecentvec _{\Ysswayindexother}}

\newcommand{\ysssrcsigassocconttimecentindexotherval}
          {$\Ysssrcsigassocconttimecentindexotherval$}
\newcommand{\Ysssrcsigassocconttimecentindexotherval}
           {\Ysssrcsigassocconttimecentindexother (\Yssconttimeval)}

\newcommand{\ysssrcsigassocconttimecentindexthird}
          {$\Ysssrcsigassocconttimecentindexthird$}
\newcommand{\Ysssrcsigassocconttimecentindexthird}
           {\Ysssrcsigassocconttimecentvec _{\Ysswayindexthird}}

\newcommand{\ysssrcsigassocconttimecentindexthirdval}
          {$\Ysssrcsigassocconttimecentindexthirdval$}
\newcommand{\Ysssrcsigassocconttimecentindexthirdval}
           {\Ysssrcsigassocconttimecentindexthird (\Yssconttimeval)}

\newcommand{\ysssrcsigassocconttimecentindexfourth}
{$\Ysssrcsigassocconttimecentindexfourth$}
\newcommand{\Ysssrcsigassocconttimecentindexfourth}
{\Ysssrcsigassocconttimecentvec _{\Ysswayindexfourth}}

\newcommand{\ysssrcsigassocconttimecentindexfourthval}
{$\Ysssrcsigassocconttimecentindexfourthval$}
\newcommand{\Ysssrcsigassocconttimecentindexfourthval}
{\Ysssrcsigassocconttimecentindexfourth (\Yssconttimeval)}

\newcommand{\ysssrcsigassocconttimecentlast}
          {$\Ysssrcsigassocconttimecentlast$}
\newcommand{\Ysssrcsigassocconttimecentlast}
           {\Ysssrcsigassocconttimecentvec _{\Ysssrcnb}}

\newcommand{\ysssrcsigassocconttimecentlastval}
          {$\Ysssrcsigassocconttimecentlastval$}
\newcommand{\Ysssrcsigassocconttimecentlastval}
           {\Ysssrcsigassocconttimecentlast (\Yssconttimeval)}


\newcommand{\ysssrcsigconttimecontfreqnoncentvec}
          {$\Ysssrcsigconttimecontfreqnoncentvec$}
\newcommand{\Ysssrcsigconttimecontfreqnoncentvec}
           {S}

\newcommand{\ysssrcsigconttimecontfreqnoncentvecval}
          {$\Ysssrcsigconttimecontfreqnoncentvecval$}
\newcommand{\Ysssrcsigconttimecontfreqnoncentvecval}
           {\Ysssrcsigconttimecontfreqnoncentvec ( \Yssconttimeval , \Ysscontfreqval ) }

\newcommand{\ysssrcsigconttimecontfreqnoncentone}
          {$\Ysssrcsigconttimecontfreqnoncentone$}
\newcommand{\Ysssrcsigconttimecontfreqnoncentone}
           {\Ysssrcsigconttimecontfreqnoncentvec _{1}}

\newcommand{\ysssrcsigconttimecontfreqnoncentoneval}
          {$\Ysssrcsigconttimecontfreqnoncentoneval$}
\newcommand{\Ysssrcsigconttimecontfreqnoncentoneval}
           {\Ysssrcsigconttimecontfreqnoncentone ( \Yssconttimeval , \Ysscontfreqval ) }

\newcommand{\ysssrcsigconttimecontfreqnoncenttwo}
          {$\Ysssrcsigconttimecontfreqnoncenttwo$}
\newcommand{\Ysssrcsigconttimecontfreqnoncenttwo}
           {\Ysssrcsigconttimecontfreqnoncentvec _{2}}

\newcommand{\ysssrcsigconttimecontfreqnoncenttwoval}
          {$\Ysssrcsigconttimecontfreqnoncenttwoval$}
\newcommand{\Ysssrcsigconttimecontfreqnoncenttwoval}
           {\Ysssrcsigconttimecontfreqnoncenttwo ( \Yssconttimeval , \Ysscontfreqval ) }

\newcommand{\ysssrcsigconttimecontfreqnoncentindex}
          {$\Ysssrcsigconttimecontfreqnoncentindex$}
\newcommand{\Ysssrcsigconttimecontfreqnoncentindex}
           {\Ysssrcsigconttimecontfreqnoncentvec _{\Ysswayindex}}

\newcommand{\ysssrcsigconttimecontfreqnoncentindexval}
          {$\Ysssrcsigconttimecontfreqnoncentindexval$}
\newcommand{\Ysssrcsigconttimecontfreqnoncentindexval}
           {\Ysssrcsigconttimecontfreqnoncentindex ( \Yssconttimeval , \Ysscontfreqval ) }

\newcommand{\ysssrcsigconttimecontfreqnoncentindexother}
          {$\Ysssrcsigconttimecontfreqnoncentindexother$}
\newcommand{\Ysssrcsigconttimecontfreqnoncentindexother}
           {\Ysssrcsigconttimecontfreqnoncentvec _{\Ysswayindexother}}

\newcommand{\ysssrcsigconttimecontfreqnoncentindexotherval}
          {$\Ysssrcsigconttimecontfreqnoncentindexotherval$}
\newcommand{\Ysssrcsigconttimecontfreqnoncentindexotherval}
           {\Ysssrcsigconttimecontfreqnoncentindexother ( \Yssconttimeval , \Ysscontfreqval ) }

\newcommand{\ysssrcsigconttimecontfreqnoncentindexthird}
          {$\Ysssrcsigconttimecontfreqnoncentindexthird$}
\newcommand{\Ysssrcsigconttimecontfreqnoncentindexthird}
           {\Ysssrcsigconttimecontfreqnoncentvec _{\Ysswayindexthird}}

\newcommand{\ysssrcsigconttimecontfreqnoncentindexthirdval}
          {$\Ysssrcsigconttimecontfreqnoncentindexthirdval$}
\newcommand{\Ysssrcsigconttimecontfreqnoncentindexthirdval}
           {\Ysssrcsigconttimecontfreqnoncentindexthird ( \Yssconttimeval , \Ysscontfreqval ) }

\newcommand{\ysssrcsigconttimecontfreqnoncentindexfourth}
          {$\Ysssrcsigconttimecontfreqnoncentindexfourth$}
\newcommand{\Ysssrcsigconttimecontfreqnoncentindexfourth}
           {\Ysssrcsigconttimecontfreqnoncentvec _{\Ysswayindexfourth}}

\newcommand{\ysssrcsigconttimecontfreqnoncentindexfourthval}
          {$\Ysssrcsigconttimecontfreqnoncentindexfourthval$}
\newcommand{\Ysssrcsigconttimecontfreqnoncentindexfourthval}
           {\Ysssrcsigconttimecontfreqnoncentindexfourth ( \Yssconttimeval , \Ysscontfreqval ) }


\newcommand{\ysssrcsigconttimecontscalecentindexother}
          {$\Ysssrcsigconttimecontscalecentindexother$}
\newcommand{\Ysssrcsigconttimecontscalecentindexother}
           {\Ysstimescalecontcoefnot _{\Ysssrcsigconttimecentvec _{\Ysswayindexother}}}

\newcommand{\ysssrcsigconttimecontscalecentindexotherval}
          {$\Ysssrcsigconttimecontscalecentindexotherval$}
\newcommand{\Ysssrcsigconttimecontscalecentindexotherval}
           {\Ysssrcsigconttimecontscalecentindexother
        ( \Ysstimescalecontshiftval , \Ysstimescalecontscaleval )
           }

\newcommand{\ysssrcsigconttimecontscalecentindexothervalsubstd}
          {$\Ysssrcsigconttimecontscalecentindexothervalsubstd$}
\newcommand{\Ysssrcsigconttimecontscalecentindexothervalsubstd}
           {\Ysssrcsigconttimecontscalecentindexother
        ( \Ysstimescalecontshiftval , \Ysstimescalecontscalevalsubstd )
           }


\newcommand{\ysssrcsigdisctimecentvec}
          {$\Ysssrcsigdisctimecentvec$}
\newcommand{\Ysssrcsigdisctimecentvec}
           {s}

\newcommand{\ysssrcsigdisctimecentvecval}
          {$\Ysssrcsigdisctimecentvecval$}
\newcommand{\Ysssrcsigdisctimecentvecval}
           {\Ysssrcsigdisctimecentvec (\Yssdisctimeval)}

\newcommand{\ysssrcsigdisctimecentone}
          {$\Ysssrcsigdisctimecentone$}
\newcommand{\Ysssrcsigdisctimecentone}
           {\Ysssrcsigdisctimecentvec _{1}}

\newcommand{\ysssrcsigdisctimecentoneval}
          {$\Ysssrcsigdisctimecentoneval$}
\newcommand{\Ysssrcsigdisctimecentoneval}
           {\Ysssrcsigdisctimecentone (\Yssdisctimeval)}

\newcommand{\ysssrcsigdisctimecenttwo}
          {$\Ysssrcsigdisctimecenttwo$}
\newcommand{\Ysssrcsigdisctimecenttwo}
           {\Ysssrcsigdisctimecentvec _{2}}

\newcommand{\ysssrcsigdisctimecenttwoval}
          {$\Ysssrcsigdisctimecenttwoval$}
\newcommand{\Ysssrcsigdisctimecenttwoval}
           {\Ysssrcsigdisctimecenttwo (\Yssdisctimeval)}

\newcommand{\ysssrcsigdisctimecentthree}
          {$\Ysssrcsigdisctimecentthree$}
\newcommand{\Ysssrcsigdisctimecentthree}
           {\Ysssrcsigdisctimecentvec _{3}}

\newcommand{\ysssrcsigdisctimecentthreeval}
          {$\Ysssrcsigdisctimecentthreeval$}
\newcommand{\Ysssrcsigdisctimecentthreeval}
           {\Ysssrcsigdisctimecentthree (\Yssdisctimeval)}

\newcommand{\ysssrcsigdisctimecentindex}
          {$\Ysssrcsigdisctimecentindex$}
\newcommand{\Ysssrcsigdisctimecentindex}
           {\Ysssrcsigdisctimecentvec _{\Ysswayindex}}

\newcommand{\ysssrcsigdisctimecentindexval}
          {$\Ysssrcsigdisctimecentindexval$}
\newcommand{\Ysssrcsigdisctimecentindexval}
           {\Ysssrcsigdisctimecentindex (\Yssdisctimeval)}

\newcommand{\ysssrcsigdisctimecentindexother}
          {$\Ysssrcsigdisctimecentindexother$}
\newcommand{\Ysssrcsigdisctimecentindexother}
           {\Ysssrcsigdisctimecentvec _{\Ysswayindexother}}

\newcommand{\ysssrcsigdisctimecentindexotherval}
          {$\Ysssrcsigdisctimecentindexotherval$}
\newcommand{\Ysssrcsigdisctimecentindexotherval}
           {\Ysssrcsigdisctimecentindexother (\Yssdisctimeval)}

\newcommand{\ysssrcsigdisctimecentindexthird}
          {$\Ysssrcsigdisctimecentindexthird$}
\newcommand{\Ysssrcsigdisctimecentindexthird}
           {\Ysssrcsigdisctimecentvec _{\Ysswayindexthird}}

\newcommand{\ysssrcsigdisctimecentindexthirdval}
          {$\Ysssrcsigdisctimecentindexthirdval$}
\newcommand{\Ysssrcsigdisctimecentindexthirdval}
           {\Ysssrcsigdisctimecentindexthird (\Yssdisctimeval)}

\newcommand{\ysssrcsigdisctimecentlast}
          {$\Ysssrcsigdisctimecentlast$}
\newcommand{\Ysssrcsigdisctimecentlast}
           {\Ysssrcsigdisctimecentvec _{\Ysssrcnb}}

\newcommand{\ysssrcsigdisctimecentlastval}
          {$\Ysssrcsigdisctimecentlastval$}
\newcommand{\Ysssrcsigdisctimecentlastval}
           {\Ysssrcsigdisctimecentlast (\Yssdisctimeval)}


\newcommand{\ysssrcsigassoctwodisctimecentvec}
          {$\Ysssrcsigassoctwodisctimecentvec$}
\newcommand{\Ysssrcsigassoctwodisctimecentvec}
           {\tilde{\Ysssrcsigdisctimecentvec}}

\newcommand{\ysssrcsigassoctwodisctimecentvecval}
          {$\Ysssrcsigassoctwodisctimecentvecval$}
\newcommand{\Ysssrcsigassoctwodisctimecentvecval}
           {\Ysssrcsigassoctwodisctimecentvec (\Yssdisctimeval)}

\newcommand{\ysssrcsigassoctwodisctimecentone}
{$\Ysssrcsigassoctwodisctimecentone$}
\newcommand{\Ysssrcsigassoctwodisctimecentone}
{\Ysssrcsigassoctwodisctimecentvec _{1}}

\newcommand{\ysssrcsigassoctwodisctimecentoneval}
{$\Ysssrcsigassoctwodisctimecentoneval$}
\newcommand{\Ysssrcsigassoctwodisctimecentoneval}
{\Ysssrcsigassoctwodisctimecentone (\Yssdisctimeval)}

\newcommand{\ysssrcsigassoctwodisctimecenttwo}
{$\Ysssrcsigassoctwodisctimecenttwo$}
\newcommand{\Ysssrcsigassoctwodisctimecenttwo}
{\Ysssrcsigassoctwodisctimecentvec _{2}}

\newcommand{\ysssrcsigassoctwodisctimecenttwoval}
{$\Ysssrcsigassoctwodisctimecenttwoval$}
\newcommand{\Ysssrcsigassoctwodisctimecenttwoval}
{\Ysssrcsigassoctwodisctimecenttwo (\Yssdisctimeval)}

\newcommand{\ysssrcsigassoctwodisctimecentindex}
{$\Ysssrcsigassoctwodisctimecentindex$}
\newcommand{\Ysssrcsigassoctwodisctimecentindex}
{\Ysssrcsigassoctwodisctimecentvec _{\Ysswayindex}}

\newcommand{\ysssrcsigassoctwodisctimecentindexval}
{$\Ysssrcsigassoctwodisctimecentindexval$}
\newcommand{\Ysssrcsigassoctwodisctimecentindexval}
{\Ysssrcsigassoctwodisctimecentindex (\Yssdisctimeval)}

\newcommand{\ysssrcsigassoctwodisctimecentindexother}
          {$\Ysssrcsigassoctwodisctimecentindexother$}
\newcommand{\Ysssrcsigassoctwodisctimecentindexother}
           {\Ysssrcsigassoctwodisctimecentvec _{\Ysswayindexother}}

\newcommand{\ysssrcsigassoctwodisctimecentindexotherval}
          {$\Ysssrcsigassoctwodisctimecentindexotherval$}
\newcommand{\Ysssrcsigassoctwodisctimecentindexotherval}
           {\Ysssrcsigassoctwodisctimecentindexother (\Yssdisctimeval)}


\newcommand{\ysssrcsigdisctimecontfreqnotonenoncentvec}
          {$\Ysssrcsigdisctimecontfreqnotonenoncentvec$}
\newcommand{\Ysssrcsigdisctimecontfreqnotonenoncentvec}
           {S}

\newcommand{\ysssrcsigdisctimecontfreqnotonenoncentvecval}
          {$\Ysssrcsigdisctimecontfreqnotonenoncentvecval$}
\newcommand{\Ysssrcsigdisctimecontfreqnotonenoncentvecval}
           {\Ysssrcsigdisctimecontfreqnotonenoncentvec ( \Ysscontfreqval ) }

\newcommand{\ysssrcsigdisctimecontfreqnotonenoncentone}
          {$\Ysssrcsigdisctimecontfreqnotonenoncentone$}
\newcommand{\Ysssrcsigdisctimecontfreqnotonenoncentone}
           {\Ysssrcsigdisctimecontfreqnotonenoncentvec _{1}}

\newcommand{\ysssrcsigdisctimecontfreqnotonenoncentoneval}
          {$\Ysssrcsigdisctimecontfreqnotonenoncentoneval$}
\newcommand{\Ysssrcsigdisctimecontfreqnotonenoncentoneval}
           {\Ysssrcsigdisctimecontfreqnotonenoncentone (\Ysscontfreqval ) }

\newcommand{\ysssrcsigdisctimecontfreqnotonenoncentindexother}
          {$\Ysssrcsigdisctimecontfreqnotonenoncentindexother$}
\newcommand{\Ysssrcsigdisctimecontfreqnotonenoncentindexother}
           {\Ysssrcsigdisctimecontfreqnotonenoncentvec _{\Ysswayindexother}}

\newcommand{\ysssrcsigdisctimecontfreqnotonenoncentindexotherval}
          {$\Ysssrcsigdisctimecontfreqnotonenoncentindexotherval$}
\newcommand{\Ysssrcsigdisctimecontfreqnotonenoncentindexotherval}
           {\Ysssrcsigdisctimecontfreqnotonenoncentindexother (\Ysscontfreqval ) }

\newcommand{\ysssrcsigdisctimecontfreqnotonenoncentlast}
          {$\Ysssrcsigdisctimecontfreqnotonenoncentlast$}
\newcommand{\Ysssrcsigdisctimecontfreqnotonenoncentlast}
           {\Ysssrcsigdisctimecontfreqnotonenoncentvec _{\Ysssrcnb}}

\newcommand{\ysssrcsigdisctimecontfreqnotonenoncentlastval}
          {$\Ysssrcsigdisctimecontfreqnotonenoncentlastval$}
\newcommand{\Ysssrcsigdisctimecontfreqnotonenoncentlastval}
           {\Ysssrcsigdisctimecontfreqnotonenoncentlast (\Ysscontfreqval ) }


\newcommand{\ysssrcsigassocdisctimecontfreqnotonenoncentvec}
          {$\Ysssrcsigassocdisctimecontfreqnotonenoncentvec$}
\newcommand{\Ysssrcsigassocdisctimecontfreqnotonenoncentvec}
           {\Ysssrcsigdisctimecontfreqnotonenoncentvec ^{\prime}}

\newcommand{\ysssrcsigassocdisctimecontfreqnotonenoncentvecval}
          {$\Ysssrcsigassocdisctimecontfreqnotonenoncentvecval$}
\newcommand{\Ysssrcsigassocdisctimecontfreqnotonenoncentvecval}
           {\Ysssrcsigassocdisctimecontfreqnotonenoncentvec ( \Ysscontfreqval ) }

\newcommand{\ysssrcsigassocdisctimecontfreqnotonenoncentone}
          {$\Ysssrcsigassocdisctimecontfreqnotonenoncentone$}
\newcommand{\Ysssrcsigassocdisctimecontfreqnotonenoncentone}
           {\Ysssrcsigassocdisctimecontfreqnotonenoncentvec _{1}}

\newcommand{\ysssrcsigassocdisctimecontfreqnotonenoncentoneval}
          {$\Ysssrcsigassocdisctimecontfreqnotonenoncentoneval$}
\newcommand{\Ysssrcsigassocdisctimecontfreqnotonenoncentoneval}
           {\Ysssrcsigassocdisctimecontfreqnotonenoncentone (\Ysscontfreqval ) }

\newcommand{\ysssrcsigassocdisctimecontfreqnotonenoncentindexother}
          {$\Ysssrcsigassocdisctimecontfreqnotonenoncentindexother$}
\newcommand{\Ysssrcsigassocdisctimecontfreqnotonenoncentindexother}
           {\Ysssrcsigassocdisctimecontfreqnotonenoncentvec _{\Ysswayindexother}}

\newcommand{\ysssrcsigassocdisctimecontfreqnotonenoncentindexotherval}
          {$\Ysssrcsigassocdisctimecontfreqnotonenoncentindexotherval$}
\newcommand{\Ysssrcsigassocdisctimecontfreqnotonenoncentindexotherval}
           {\Ysssrcsigassocdisctimecontfreqnotonenoncentindexother (\Ysscontfreqval ) }

\newcommand{\ysssrcsigassocdisctimecontfreqnotonenoncentlast}
          {$\Ysssrcsigassocdisctimecontfreqnotonenoncentlast$}
\newcommand{\Ysssrcsigassocdisctimecontfreqnotonenoncentlast}
           {\Ysssrcsigassocdisctimecontfreqnotonenoncentvec _{\Ysssrcnb}}

\newcommand{\ysssrcsigassocdisctimecontfreqnotonenoncentlastval}
          {$\Ysssrcsigassocdisctimecontfreqnotonenoncentlastval$}
\newcommand{\Ysssrcsigassocdisctimecontfreqnotonenoncentlastval}
           {\Ysssrcsigassocdisctimecontfreqnotonenoncentlast (\Ysscontfreqval ) }


\newcommand{\ysssrcsigdisctimecontfreqnoncentvec}
          {$\Ysssrcsigdisctimecontfreqnoncentvec$}
\newcommand{\Ysssrcsigdisctimecontfreqnoncentvec}
           {S}

\newcommand{\ysssrcsigdisctimecontfreqnoncentvecval}
          {$\Ysssrcsigdisctimecontfreqnoncentvecval$}
\newcommand{\Ysssrcsigdisctimecontfreqnoncentvecval}
           {\Ysssrcsigdisctimecontfreqnoncentvec ( \Yssdisctimeval , \Ysscontfreqval ) }

\newcommand{\ysssrcsigdisctimecontfreqnoncentone}
          {$\Ysssrcsigdisctimecontfreqnoncentone$}
\newcommand{\Ysssrcsigdisctimecontfreqnoncentone}
           {\Ysssrcsigdisctimecontfreqnoncentvec _{1}}

\newcommand{\ysssrcsigdisctimecontfreqnoncentoneval}
          {$\Ysssrcsigdisctimecontfreqnoncentoneval$}
\newcommand{\Ysssrcsigdisctimecontfreqnoncentoneval}
           {\Ysssrcsigdisctimecontfreqnoncentone ( \Yssdisctimeval , \Ysscontfreqval ) }

\newcommand{\ysssrcsigdisctimecontfreqnoncentindex}
          {$\Ysssrcsigdisctimecontfreqnoncentindex$}
\newcommand{\Ysssrcsigdisctimecontfreqnoncentindex}
           {\Ysssrcsigdisctimecontfreqnoncentvec _{\Ysswayindex}}

\newcommand{\ysssrcsigdisctimecontfreqnoncentindexval}
          {$\Ysssrcsigdisctimecontfreqnoncentindexval$}
\newcommand{\Ysssrcsigdisctimecontfreqnoncentindexval}
           {\Ysssrcsigdisctimecontfreqnoncentindex ( \Yssdisctimeval , \Ysscontfreqval ) }

\newcommand{\ysssrcsigdisctimecontfreqnoncentindexother}
          {$\Ysssrcsigdisctimecontfreqnoncentindexother$}
\newcommand{\Ysssrcsigdisctimecontfreqnoncentindexother}
           {\Ysssrcsigdisctimecontfreqnoncentvec _{\Ysswayindexother}}

\newcommand{\ysssrcsigdisctimecontfreqnoncentindexotherval}
          {$\Ysssrcsigdisctimecontfreqnoncentindexotherval$}
\newcommand{\Ysssrcsigdisctimecontfreqnoncentindexotherval}
           {\Ysssrcsigdisctimecontfreqnoncentindexother ( \Yssdisctimeval , \Ysscontfreqval ) }

\newcommand{\ysssrcsigdisctimecontfreqnoncentindexthird}
          {$\Ysssrcsigdisctimecontfreqnoncentindexthird$}
\newcommand{\Ysssrcsigdisctimecontfreqnoncentindexthird}
           {\Ysssrcsigdisctimecontfreqnoncentvec _{\Ysswayindexthird}}

\newcommand{\ysssrcsigdisctimecontfreqnoncentindexthirdval}
          {$\Ysssrcsigdisctimecontfreqnoncentindexthirdval$}
\newcommand{\Ysssrcsigdisctimecontfreqnoncentindexthirdval}
           {\Ysssrcsigdisctimecontfreqnoncentindexthird ( \Yssdisctimeval , \Ysscontfreqval ) }


\newcommand{\ysssrcsigassoctwodisctimecontfreqnoncentvec}
          {$\Ysssrcsigassoctwodisctimecontfreqnoncentvec$}
\newcommand{\Ysssrcsigassoctwodisctimecontfreqnoncentvec}
           {\tilde{\Ysssrcsigdisctimecontfreqnoncentvec}}

\newcommand{\ysssrcsigassoctwodisctimecontfreqnoncentvecval}
          {$\Ysssrcsigassoctwodisctimecontfreqnoncentvecval$}
\newcommand{\Ysssrcsigassoctwodisctimecontfreqnoncentvecval}
           {\Ysssrcsigassoctwodisctimecontfreqnoncentvec ( \Yssdisctimeval , \Ysscontfreqval ) }

\newcommand{\ysssrcsigassoctwodisctimecontfreqnoncentindexother}
          {$\Ysssrcsigassoctwodisctimecontfreqnoncentindexother$}
\newcommand{\Ysssrcsigassoctwodisctimecontfreqnoncentindexother}
           {\Ysssrcsigassoctwodisctimecontfreqnoncentvec _{\Ysswayindexother}}

\newcommand{\ysssrcsigassoctwodisctimecontfreqnoncentindexotherval}
          {$\Ysssrcsigassoctwodisctimecontfreqnoncentindexotherval$}
\newcommand{\Ysssrcsigassoctwodisctimecontfreqnoncentindexotherval}
           {\Ysssrcsigassoctwodisctimecontfreqnoncentindexother ( \Yssdisctimeval , \Ysscontfreqval ) }


\newcommand{\ysssrcsigdisctimecentveczt}
          {$\Ysssrcsigdisctimecentveczt$}
\newcommand{\Ysssrcsigdisctimecentveczt}
           {S}

\newcommand{\ysssrcsigdisctimecentvecztval}
          {$\Ysssrcsigdisctimecentvecztval$}
\newcommand{\Ysssrcsigdisctimecentvecztval}
           {\Ysssrcsigdisctimecentveczt (z)}

\newcommand{\ysssrcsigdisctimecentonezt}
          {$\Ysssrcsigdisctimecentonezt$}
\newcommand{\Ysssrcsigdisctimecentonezt}
           {\Ysssrcsigdisctimecentveczt _{1}}

\newcommand{\ysssrcsigdisctimecentoneztval}
          {$\Ysssrcsigdisctimecentoneztval$}
\newcommand{\Ysssrcsigdisctimecentoneztval}
           {\Ysssrcsigdisctimecentonezt (z)}

\newcommand{\ysssrcsigdisctimecenttwozt}
          {$\Ysssrcsigdisctimecenttwozt$}
\newcommand{\Ysssrcsigdisctimecenttwozt}
           {\Ysssrcsigdisctimecentveczt _{2}}

\newcommand{\ysssrcsigdisctimecenttwoztval}
          {$\Ysssrcsigdisctimecenttwoztval$}
\newcommand{\Ysssrcsigdisctimecenttwoztval}
           {\Ysssrcsigdisctimecenttwozt (z)}

\newcommand{\ysssrcsigdisctimecentthreezt}
          {$\Ysssrcsigdisctimecentthreezt$}
\newcommand{\Ysssrcsigdisctimecentthreezt}
           {\Ysssrcsigdisctimecentveczt _{3}}

\newcommand{\ysssrcsigdisctimecentthreeztval}
          {$\Ysssrcsigdisctimecentthreeztval$}
\newcommand{\Ysssrcsigdisctimecentthreeztval}
           {\Ysssrcsigdisctimecentthreezt (z)}

\newcommand{\ysssrcsigdisctimecentindexzt}
          {$\Ysssrcsigdisctimecentindexzt$}
\newcommand{\Ysssrcsigdisctimecentindexzt}
           {\Ysssrcsigdisctimecentveczt _{\Ysswayindex}}

\newcommand{\ysssrcsigdisctimecentindexztval}
          {$\Ysssrcsigdisctimecentindexztval$}
\newcommand{\Ysssrcsigdisctimecentindexztval}
           {\Ysssrcsigdisctimecentindexzt (z)}

\newcommand{\ysssrcsigdisctimecentindexotherzt}
          {$\Ysssrcsigdisctimecentindexotherzt$}
\newcommand{\Ysssrcsigdisctimecentindexotherzt}
           {\Ysssrcsigdisctimecentveczt _{\Ysswayindexother}}

\newcommand{\ysssrcsigdisctimecentindexotherztval}
          {$\Ysssrcsigdisctimecentindexotherztval$}
\newcommand{\Ysssrcsigdisctimecentindexotherztval}
           {\Ysssrcsigdisctimecentindexotherzt (z)}

\newcommand{\ysssrcsigdisctimecentlastzt}
          {$\Ysssrcsigdisctimecentlastzt$}
\newcommand{\Ysssrcsigdisctimecentlastzt}
           {\Ysssrcsigdisctimecentveczt _{\Ysssrcnb}}

\newcommand{\ysssrcsigdisctimecentlastztval}
          {$\Ysssrcsigdisctimecentlastztval$}
\newcommand{\Ysssrcsigdisctimecentlastztval}
           {\Ysssrcsigdisctimecentlastzt (z)}



\newcommand{\yssmixsigconttimecentvec}
          {$\Yssmixsigconttimecentvec$}
\newcommand{\Yssmixsigconttimecentvec}
           {x}

\newcommand{\yssmixsigconttimecentvecval}
          {$\Yssmixsigconttimecentvecval$}
\newcommand{\Yssmixsigconttimecentvecval}
           {\Yssmixsigconttimecentvec (\Yssconttimeval)}

\newcommand{\yssmixsigconttimecentone}
          {$\Yssmixsigconttimecentone$}
\newcommand{\Yssmixsigconttimecentone}
           {\Yssmixsigconttimecentvec _{1}}

\newcommand{\yssmixsigconttimecentoneval}
          {$\Yssmixsigconttimecentoneval$}
\newcommand{\Yssmixsigconttimecentoneval}
           {\Yssmixsigconttimecentone (\Yssconttimeval)}

\newcommand{\yssmixsigconttimecenttwo}
          {$\Yssmixsigconttimecenttwo$}
\newcommand{\Yssmixsigconttimecenttwo}
           {\Yssmixsigconttimecentvec _{2}}

\newcommand{\yssmixsigconttimecenttwoval}
          {$\Yssmixsigconttimecenttwoval$}
\newcommand{\Yssmixsigconttimecenttwoval}
           {\Yssmixsigconttimecenttwo (\Yssconttimeval)}

\newcommand{\yssmixsigconttimecentindex}
          {$\Yssmixsigconttimecentindex$}
\newcommand{\Yssmixsigconttimecentindex}
           {\Yssmixsigconttimecentvec _{\Ysswayindex}}

\newcommand{\yssmixsigconttimecentindexval}
          {$\Yssmixsigconttimecentindexval$}
\newcommand{\Yssmixsigconttimecentindexval}
           {\Yssmixsigconttimecentindex (\Yssconttimeval)}

\newcommand{\yssmixsigconttimecentindexother}
          {$\Yssmixsigconttimecentindexother$}
\newcommand{\Yssmixsigconttimecentindexother}
           {\Yssmixsigconttimecentvec _{\Ysswayindexother}}

\newcommand{\yssmixsigconttimecentindexotherval}
          {$\Yssmixsigconttimecentindexotherval$}
\newcommand{\Yssmixsigconttimecentindexotherval}
           {\Yssmixsigconttimecentindexother (\Yssconttimeval)}

\newcommand{\yssmixsigconttimecentindexthird}
          {$\Yssmixsigconttimecentindexthird$}
\newcommand{\Yssmixsigconttimecentindexthird}
           {\Yssmixsigconttimecentvec _{\Ysswayindexthird}}

\newcommand{\yssmixsigconttimecentindexthirdval}
          {$\Yssmixsigconttimecentindexthirdval$}
\newcommand{\Yssmixsigconttimecentindexthirdval}
           {\Yssmixsigconttimecentindexthird (\Yssconttimeval)}

\newcommand{\yssmixsigconttimecentlast}
          {$\Yssmixsigconttimecentlast$}
\newcommand{\Yssmixsigconttimecentlast}
           {\Yssmixsigconttimecentvec _{\Ysssensnb}}

\newcommand{\yssmixsigconttimecentlastval}
          {$\Yssmixsigconttimecentlastval$}
\newcommand{\Yssmixsigconttimecentlastval}
           {\Yssmixsigconttimecentlast (\Yssconttimeval)}

\newcommand{\yssmixsigconttimecentsrcnb}
          {$\Yssmixsigconttimecentsrcnb$}
\newcommand{\Yssmixsigconttimecentsrcnb}
           {\Yssmixsigconttimecentvec _{\Ysssrcnb}}


\newcommand{\yssmixsigconttimecentsrcnbval}
          {$\Yssmixsigconttimecentsrcnbval$}
\newcommand{\Yssmixsigconttimecentsrcnbval}
           {\Yssmixsigconttimecentsrcnb (\Yssconttimeval)}


\newcommand{\yssmixsigassocconttimecentvec}
{$\Yssmixsigassocconttimecentvec$}
\newcommand{\Yssmixsigassocconttimecentvec}
{\Yssmixsigconttimecentvec ^{\prime}}

\newcommand{\yssmixsigassocconttimecentvecval}
{$\Yssmixsigassocconttimecentvecval$}
\newcommand{\Yssmixsigassocconttimecentvecval}
{\Yssmixsigassocconttimecentvec (\Yssconttimeval)}

\newcommand{\yssmixsigassocconttimecentone}
{$\Yssmixsigassocconttimecentone$}
\newcommand{\Yssmixsigassocconttimecentone}
{\Yssmixsigassocconttimecentvec _{1}}

\newcommand{\yssmixsigassocconttimecentoneval}
{$\Yssmixsigassocconttimecentoneval$}
\newcommand{\Yssmixsigassocconttimecentoneval}
{\Yssmixsigassocconttimecentone (\Yssconttimeval)}

\newcommand{\yssmixsigassocconttimecentindex}
{$\Yssmixsigassocconttimecentindex$}
\newcommand{\Yssmixsigassocconttimecentindex}
{\Yssmixsigassocconttimecentvec _{\Ysswayindex}}

\newcommand{\yssmixsigassocconttimecentindexval}
{$\Yssmixsigassocconttimecentindexval$}
\newcommand{\Yssmixsigassocconttimecentindexval}
{\Yssmixsigassocconttimecentindex (\Yssconttimeval)}

\newcommand{\yssmixsigassocconttimecentindexthird}
{$\Yssmixsigassocconttimecentindexthird$}
\newcommand{\Yssmixsigassocconttimecentindexthird}
{\Yssmixsigassocconttimecentvec _{\Ysswayindexthird}}

\newcommand{\yssmixsigassocconttimecentindexthirdval}
{$\Yssmixsigassocconttimecentindexthirdval$}
\newcommand{\Yssmixsigassocconttimecentindexthirdval}
{\Yssmixsigassocconttimecentindexthird (\Yssconttimeval)}


\newcommand{\yssmixsigconttimecontfreqnoncentvec}
          {$\Yssmixsigconttimecontfreqnoncentvec$}
\newcommand{\Yssmixsigconttimecontfreqnoncentvec}
           {X}

\newcommand{\yssmixsigconttimecontfreqnoncentvecval}
          {$\Yssmixsigconttimecontfreqnoncentvecval$}
\newcommand{\Yssmixsigconttimecontfreqnoncentvecval}
           {\Yssmixsigconttimecontfreqnoncentvec ( \Yssconttimeval , \Ysscontfreqval ) }

\newcommand{\yssmixsigconttimecontfreqnoncentone}
          {$\Yssmixsigconttimecontfreqnoncentone$}
\newcommand{\Yssmixsigconttimecontfreqnoncentone}
           {\Yssmixsigconttimecontfreqnoncentvec _{1}}

\newcommand{\yssmixsigconttimecontfreqnoncentoneval}
          {$\Yssmixsigconttimecontfreqnoncentoneval$}
\newcommand{\Yssmixsigconttimecontfreqnoncentoneval}
           {\Yssmixsigconttimecontfreqnoncentone ( \Yssconttimeval , \Ysscontfreqval ) }

\newcommand{\yssmixsigconttimecontfreqnoncenttwo}
          {$\Yssmixsigconttimecontfreqnoncenttwo$}
\newcommand{\Yssmixsigconttimecontfreqnoncenttwo}
           {\Yssmixsigconttimecontfreqnoncentvec _{2}}

\newcommand{\yssmixsigconttimecontfreqnoncenttwoval}
          {$\Yssmixsigconttimecontfreqnoncenttwoval$}
\newcommand{\Yssmixsigconttimecontfreqnoncenttwoval}
           {\Yssmixsigconttimecontfreqnoncenttwo ( \Yssconttimeval , \Ysscontfreqval ) }

\newcommand{\yssmixsigconttimecontfreqnoncentindex}
          {$\Yssmixsigconttimecontfreqnoncentindex$}
\newcommand{\Yssmixsigconttimecontfreqnoncentindex}
           {\Yssmixsigconttimecontfreqnoncentvec _{\Ysswayindex}}

\newcommand{\yssmixsigconttimecontfreqnoncentindexval}
          {$\Yssmixsigconttimecontfreqnoncentindexval$}
\newcommand{\Yssmixsigconttimecontfreqnoncentindexval}
           {\Yssmixsigconttimecontfreqnoncentindex ( \Yssconttimeval , \Ysscontfreqval ) }

\newcommand{\yssmixsigconttimecontfreqnoncentindexother}
          {$\Yssmixsigconttimecontfreqnoncentindexother$}
\newcommand{\Yssmixsigconttimecontfreqnoncentindexother}
           {\Yssmixsigconttimecontfreqnoncentvec _{\Ysswayindexother}}

\newcommand{\yssmixsigconttimecontfreqnoncentindexotherval}
          {$\Yssmixsigconttimecontfreqnoncentindexotherval$}
\newcommand{\Yssmixsigconttimecontfreqnoncentindexotherval}
           {\Yssmixsigconttimecontfreqnoncentindexother ( \Yssconttimeval , \Ysscontfreqval ) }

\newcommand{\yssmixsigconttimecontfreqnoncentindexthird}
          {$\Yssmixsigconttimecontfreqnoncentindexthird$}
\newcommand{\Yssmixsigconttimecontfreqnoncentindexthird}
           {\Yssmixsigconttimecontfreqnoncentvec _{\Ysswayindexthird}}

\newcommand{\yssmixsigconttimecontfreqnoncentindexthirdval}
          {$\Yssmixsigconttimecontfreqnoncentindexthirdval$}
\newcommand{\Yssmixsigconttimecontfreqnoncentindexthirdval}
           {\Yssmixsigconttimecontfreqnoncentindexthird ( \Yssconttimeval , \Ysscontfreqval ) }


\newcommand{\yssmixsigconttimecontscalecentindex}
          {$\Yssmixsigconttimecontscalecentindex$}
\newcommand{\Yssmixsigconttimecontscalecentindex}
           {\Ysstimescalecontcoefnot _{\Yssmixsigconttimecentvec _{\Ysswayindex}}}

\newcommand{\yssmixsigconttimecontscalecentindexval}
          {$\Yssmixsigconttimecontscalecentindexval$}
\newcommand{\Yssmixsigconttimecontscalecentindexval}
           {\Yssmixsigconttimecontscalecentindex
        ( \Ysstimescalecontshiftval , \Ysstimescalecontscaleval )
           }

\newcommand{\yssmixsigconttimecontscalecentindexvalsubstd}
          {$\Yssmixsigconttimecontscalecentindexvalsubstd$}
\newcommand{\Yssmixsigconttimecontscalecentindexvalsubstd}
           {\Yssmixsigconttimecontscalecentindex
        ( \Ysstimescalecontshiftval , \Ysstimescalecontscalevalsubstd )
           }


\newcommand{\yssmixsigdisctimecentvec}
          {$\Yssmixsigdisctimecentvec$}
\newcommand{\Yssmixsigdisctimecentvec}
           {x}

\newcommand{\yssmixsigdisctimecentvecval}
          {$\Yssmixsigdisctimecentvecval$}
\newcommand{\Yssmixsigdisctimecentvecval}
           {\Yssmixsigdisctimecentvec (\Yssdisctimeval)}

\newcommand{\yssmixsigdisctimecentone}
          {$\Yssmixsigdisctimecentone$}
\newcommand{\Yssmixsigdisctimecentone}
           {\Yssmixsigdisctimecentvec _{1}}

\newcommand{\yssmixsigdisctimecentoneval}
          {$\Yssmixsigdisctimecentoneval$}
\newcommand{\Yssmixsigdisctimecentoneval}
           {\Yssmixsigdisctimecentone (\Yssdisctimeval)}

\newcommand{\yssmixsigdisctimecenttwo}
          {$\Yssmixsigdisctimecenttwo$}
\newcommand{\Yssmixsigdisctimecenttwo}
           {\Yssmixsigdisctimecentvec _{2}}

\newcommand{\yssmixsigdisctimecenttwoval}
          {$\Yssmixsigdisctimecenttwoval$}
\newcommand{\Yssmixsigdisctimecenttwoval}
           {\Yssmixsigdisctimecenttwo (\Yssdisctimeval)}

\newcommand{\yssmixsigdisctimecentthree}
{$\Yssmixsigdisctimecentthree$}
\newcommand{\Yssmixsigdisctimecentthree}
{\Yssmixsigdisctimecentvec _{3}}

\newcommand{\yssmixsigdisctimecentthreeval}
{$\Yssmixsigdisctimecentthreeval$}
\newcommand{\Yssmixsigdisctimecentthreeval}
{\Yssmixsigdisctimecentthree (\Yssdisctimeval)}

\newcommand{\yssmixsigdisctimecentindex}
          {$\Yssmixsigdisctimecentindex$}
\newcommand{\Yssmixsigdisctimecentindex}
           {\Yssmixsigdisctimecentvec _{\Ysswayindex}}

\newcommand{\yssmixsigdisctimecentindexval}
          {$\Yssmixsigdisctimecentindexval$}
\newcommand{\Yssmixsigdisctimecentindexval}
           {\Yssmixsigdisctimecentindex (\Yssdisctimeval)}

\newcommand{\yssmixsigdisctimecentindexother}
{$\Yssmixsigdisctimecentindexother$}
\newcommand{\Yssmixsigdisctimecentindexother}
{\Yssmixsigdisctimecentvec _{\Ysswayindexother}}

\newcommand{\yssmixsigdisctimecentindexotherval}
{$\Yssmixsigdisctimecentindexotherval$}
\newcommand{\Yssmixsigdisctimecentindexotherval}
{\Yssmixsigdisctimecentindexother (\Yssdisctimeval)}

\newcommand{\yssmixsigdisctimecentlast}
          {$\Yssmixsigdisctimecentlast$}
\newcommand{\Yssmixsigdisctimecentlast}
           {\Yssmixsigdisctimecentvec _{\Ysssensnb}}

\newcommand{\yssmixsigdisctimecentlastval}
          {$\Yssmixsigdisctimecentlastval$}
\newcommand{\Yssmixsigdisctimecentlastval}
           {\Yssmixsigdisctimecentlast (\Yssdisctimeval)}

\newcommand{\yssmixsigdisctimecentsrcnb}
          {$\Yssmixsigdisctimecentsrcnb$}
\newcommand{\Yssmixsigdisctimecentsrcnb}
           {\Yssmixsigdisctimecentvec _{\Ysssrcnb}}


\newcommand{\yssmixsigdisctimecentsrcnbval}
          {$\Yssmixsigdisctimecentsrcnbval$}
\newcommand{\Yssmixsigdisctimecentsrcnbval}
           {\Yssmixsigdisctimecentsrcnb (\Yssdisctimeval)}


\newcommand{\yssmixsigdisctimecontfreqnotonenoncentvec}
          {$\Yssmixsigdisctimecontfreqnotonenoncentvec$}
\newcommand{\Yssmixsigdisctimecontfreqnotonenoncentvec}
           {X}

\newcommand{\yssmixsigdisctimecontfreqnotonenoncentvecval}
          {$\Yssmixsigdisctimecontfreqnotonenoncentvecval$}
\newcommand{\Yssmixsigdisctimecontfreqnotonenoncentvecval}
           {\Yssmixsigdisctimecontfreqnotonenoncentvec ( \Ysscontfreqval ) }

\newcommand{\yssmixsigdisctimecontfreqnotonenoncentone}
          {$\Yssmixsigdisctimecontfreqnotonenoncentone$}
\newcommand{\Yssmixsigdisctimecontfreqnotonenoncentone}
           {\Yssmixsigdisctimecontfreqnotonenoncentvec _{1}}

\newcommand{\yssmixsigdisctimecontfreqnotonenoncentoneval}
          {$\Yssmixsigdisctimecontfreqnotonenoncentoneval$}
\newcommand{\Yssmixsigdisctimecontfreqnotonenoncentoneval}
           {\Yssmixsigdisctimecontfreqnotonenoncentone (\Ysscontfreqval ) }

\newcommand{\yssmixsigdisctimecontfreqnotonenoncenttwo}
          {$\Yssmixsigdisctimecontfreqnotonenoncenttwo$}
\newcommand{\Yssmixsigdisctimecontfreqnotonenoncenttwo}
           {\Yssmixsigdisctimecontfreqnotonenoncentvec _{2}}

\newcommand{\yssmixsigdisctimecontfreqnotonenoncenttwoval}
          {$\Yssmixsigdisctimecontfreqnotonenoncenttwoval$}
\newcommand{\Yssmixsigdisctimecontfreqnotonenoncenttwoval}
           {\Yssmixsigdisctimecontfreqnotonenoncenttwo (\Ysscontfreqval ) }

\newcommand{\yssmixsigdisctimecontfreqnotonenoncentindex}
          {$\Yssmixsigdisctimecontfreqnotonenoncentindex$}
\newcommand{\Yssmixsigdisctimecontfreqnotonenoncentindex}
           {\Yssmixsigdisctimecontfreqnotonenoncentvec _{\Ysswayindex}}

\newcommand{\yssmixsigdisctimecontfreqnotonenoncentindexval}
          {$\Yssmixsigdisctimecontfreqnotonenoncentindexval$}
\newcommand{\Yssmixsigdisctimecontfreqnotonenoncentindexval}
           {\Yssmixsigdisctimecontfreqnotonenoncentindex (\Ysscontfreqval ) }

\newcommand{\yssmixsigdisctimecontfreqnotonenoncentsrcnb}
          {$\Yssmixsigdisctimecontfreqnotonenoncentsrcnb$}
\newcommand{\Yssmixsigdisctimecontfreqnotonenoncentsrcnb}
           {\Yssmixsigdisctimecontfreqnotonenoncentvec _{\Ysssrcnb}}

\newcommand{\yssmixsigdisctimecontfreqnotonenoncentsrcnbval}
          {$\Yssmixsigdisctimecontfreqnotonenoncentsrcnbval$}
\newcommand{\Yssmixsigdisctimecontfreqnotonenoncentsrcnbval}
           {\Yssmixsigdisctimecontfreqnotonenoncentsrcnb (\Ysscontfreqval ) }


\newcommand{\yssmixsigdisctimecontfreqnoncentvec}
          {$\Yssmixsigdisctimecontfreqnoncentvec$}
\newcommand{\Yssmixsigdisctimecontfreqnoncentvec}
           {X}

\newcommand{\yssmixsigdisctimecontfreqnoncentvecval}
          {$\Yssmixsigdisctimecontfreqnoncentvecval$}
\newcommand{\Yssmixsigdisctimecontfreqnoncentvecval}
           {\Yssmixsigdisctimecontfreqnoncentvec ( \Yssdisctimeval , \Ysscontfreqval ) }

\newcommand{\yssmixsigdisctimecontfreqnoncentone}
          {$\Yssmixsigdisctimecontfreqnoncentone$}
\newcommand{\Yssmixsigdisctimecontfreqnoncentone}
           {\Yssmixsigdisctimecontfreqnoncentvec _{1}}

\newcommand{\yssmixsigdisctimecontfreqnoncentoneval}
          {$\Yssmixsigdisctimecontfreqnoncentoneval$}
\newcommand{\Yssmixsigdisctimecontfreqnoncentoneval}
           {\Yssmixsigdisctimecontfreqnoncentone ( \Yssdisctimeval , \Ysscontfreqval ) }

\newcommand{\yssmixsigdisctimecontfreqnoncentindex}
          {$\Yssmixsigdisctimecontfreqnoncentindex$}
\newcommand{\Yssmixsigdisctimecontfreqnoncentindex}
           {\Yssmixsigdisctimecontfreqnoncentvec _{\Ysswayindex}}

\newcommand{\yssmixsigdisctimecontfreqnoncentindexval}
          {$\Yssmixsigdisctimecontfreqnoncentindexval$}
\newcommand{\Yssmixsigdisctimecontfreqnoncentindexval}
           {\Yssmixsigdisctimecontfreqnoncentindex ( \Yssdisctimeval , \Ysscontfreqval ) }

\newcommand{\yssmixsigdisctimecontfreqnoncentindexother}
          {$\Yssmixsigdisctimecontfreqnoncentindexother$}
\newcommand{\Yssmixsigdisctimecontfreqnoncentindexother}
           {\Yssmixsigdisctimecontfreqnoncentvec _{\Ysswayindexother}}

\newcommand{\yssmixsigdisctimecontfreqnoncentindexotherval}
          {$\Yssmixsigdisctimecontfreqnoncentindexotherval$}
\newcommand{\Yssmixsigdisctimecontfreqnoncentindexotherval}
           {\Yssmixsigdisctimecontfreqnoncentindexother ( \Yssdisctimeval , \Ysscontfreqval ) }


\newcommand{\yssmixsigdisctimecentveczt}
          {$\Yssmixsigdisctimecentveczt$}
\newcommand{\Yssmixsigdisctimecentveczt}
           {X}

\newcommand{\yssmixsigdisctimecentvecztval}
          {$\Yssmixsigdisctimecentvecztval$}
\newcommand{\Yssmixsigdisctimecentvecztval}
           {\Yssmixsigdisctimecentveczt (z)}

\newcommand{\yssmixsigdisctimecentonezt}
          {$\Yssmixsigdisctimecentonezt$}
\newcommand{\Yssmixsigdisctimecentonezt}
           {\Yssmixsigdisctimecentveczt _{1}}

\newcommand{\yssmixsigdisctimecentoneztval}
          {$\Yssmixsigdisctimecentoneztval$}
\newcommand{\Yssmixsigdisctimecentoneztval}
           {\Yssmixsigdisctimecentonezt (z)}

\newcommand{\yssmixsigdisctimecenttwozt}
          {$\Yssmixsigdisctimecenttwozt$}
\newcommand{\Yssmixsigdisctimecenttwozt}
           {\Yssmixsigdisctimecentveczt _{2}}

\newcommand{\yssmixsigdisctimecenttwoztval}
          {$\Yssmixsigdisctimecenttwoztval$}
\newcommand{\Yssmixsigdisctimecenttwoztval}
           {\Yssmixsigdisctimecenttwozt (z)}

\newcommand{\yssmixsigdisctimecentindexzt}
          {$\Yssmixsigdisctimecentindexzt$}
\newcommand{\Yssmixsigdisctimecentindexzt}
           {\Yssmixsigdisctimecentveczt _{\Ysswayindex}}

\newcommand{\yssmixsigdisctimecentindexztval}
          {$\Yssmixsigdisctimecentindexztval$}
\newcommand{\Yssmixsigdisctimecentindexztval}
           {\Yssmixsigdisctimecentindexzt (z)}

\newcommand{\yssmixsigdisctimecentlastzt}
          {$\Yssmixsigdisctimecentlastzt$}
\newcommand{\Yssmixsigdisctimecentlastzt}
           {\Yssmixsigdisctimecentveczt _{\Ysssensnb}}

\newcommand{\yssmixsigdisctimecentlastztval}
          {$\Yssmixsigdisctimecentlastztval$}
\newcommand{\Yssmixsigdisctimecentlastztval}
           {\Yssmixsigdisctimecentlastzt (z)}



\newcommand{\yssintermsepsystsigdisctimecentvec}
          {$\Yssintermsepsystsigdisctimecentvec$}
\newcommand{\Yssintermsepsystsigdisctimecentvec}
           {u}

\newcommand{\yssintermsepsystsigdisctimecentone}
          {$\Yssintermsepsystsigdisctimecentone$}
\newcommand{\Yssintermsepsystsigdisctimecentone}
           {\Yssintermsepsystsigdisctimecentvec _{1}}

\newcommand{\yssintermsepsystsigdisctimecentoneval}
          {$\Yssintermsepsystsigdisctimecentoneval$}
\newcommand{\Yssintermsepsystsigdisctimecentoneval}
           {\Yssintermsepsystsigdisctimecentone ( \Yssdisctimeval )}

\newcommand{\yssintermsepsystsigdisctimecenttwo}
          {$\Yssintermsepsystsigdisctimecenttwo$}
\newcommand{\Yssintermsepsystsigdisctimecenttwo}
           {\Yssintermsepsystsigdisctimecentvec _{2}}

\newcommand{\yssintermsepsystsigdisctimecenttwoval}
          {$\Yssintermsepsystsigdisctimecenttwoval$}
\newcommand{\Yssintermsepsystsigdisctimecenttwoval}
           {\Yssintermsepsystsigdisctimecenttwo ( \Yssdisctimeval )}

\newcommand{\yssintermsepsystsigdisctimecentindex}
          {$\Yssintermsepsystsigdisctimecentindex$}
\newcommand{\Yssintermsepsystsigdisctimecentindex}
           {\Yssintermsepsystsigdisctimecentvec _{\Ysswayindex}}

\newcommand{\yssintermsepsystsigdisctimecentindexval}
          {$\Yssintermsepsystsigdisctimecentindexval$}
\newcommand{\Yssintermsepsystsigdisctimecentindexval}
           {\Yssintermsepsystsigdisctimecentindex ( \Yssdisctimeval )}

\newcommand{\yssintermsepsystsigdisctimecentindexother}
          {$\Yssintermsepsystsigdisctimecentindexother$}
\newcommand{\Yssintermsepsystsigdisctimecentindexother}
           {\Yssintermsepsystsigdisctimecentvec _{\Ysswayindexother}}


\newcommand{\yssoutsepsystsigconttimecentvec}
          {$\Yssoutsepsystsigconttimecentvec$}
\newcommand{\Yssoutsepsystsigconttimecentvec}
           {y}

\newcommand{\yssoutsepsystsigconttimecentvecval}
          {$\Yssoutsepsystsigconttimecentvecval$}
\newcommand{\Yssoutsepsystsigconttimecentvecval}
           {\Yssoutsepsystsigconttimecentvec (\Yssconttimeval)}

\newcommand{\yssoutsepsystsigconttimecentone}
          {$\Yssoutsepsystsigconttimecentone$}
\newcommand{\Yssoutsepsystsigconttimecentone}
           {\Yssoutsepsystsigconttimecentvec _{1}}

\newcommand{\yssoutsepsystsigconttimecentoneval}
          {$\Yssoutsepsystsigconttimecentoneval$}
\newcommand{\Yssoutsepsystsigconttimecentoneval}
           {\Yssoutsepsystsigconttimecentone (\Yssconttimeval)}

\newcommand{\yssoutsepsystsigconttimecentindex}
          {$\Yssoutsepsystsigconttimecentindex$}
\newcommand{\Yssoutsepsystsigconttimecentindex}
           {\Yssoutsepsystsigconttimecentvec _{\Ysswayindex}}

\newcommand{\yssoutsepsystsigconttimecentindexval}
          {$\Yssoutsepsystsigconttimecentindexval$}
\newcommand{\Yssoutsepsystsigconttimecentindexval}
           {\Yssoutsepsystsigconttimecentindex (\Yssconttimeval)}

\newcommand{\yssoutsepsystsigconttimecentindexother}
          {$\Yssoutsepsystsigconttimecentindexother$}
\newcommand{\Yssoutsepsystsigconttimecentindexother}
           {\Yssoutsepsystsigconttimecentvec _{\Ysswayindexother}}

\newcommand{\yssoutsepsystsigconttimecentindexotherval}
          {$\Yssoutsepsystsigconttimecentindexotherval$}
\newcommand{\Yssoutsepsystsigconttimecentindexotherval}
           {\Yssoutsepsystsigconttimecentindexother (\Yssconttimeval)}

\newcommand{\yssoutsepsystsigconttimecentindexfourth}
{$\Yssoutsepsystsigconttimecentindexfourth$}
\newcommand{\Yssoutsepsystsigconttimecentindexfourth}
{\Yssoutsepsystsigconttimecentvec _{\Ysswayindexfourth}}

\newcommand{\yssoutsepsystsigconttimecentindexfourthval}
{$\Yssoutsepsystsigconttimecentindexfourthval$}
\newcommand{\Yssoutsepsystsigconttimecentindexfourthval}
{\Yssoutsepsystsigconttimecentindexfourth (\Yssconttimeval)}


\newcommand{\yssoutsepsystsigassocconttimecentvec}
          {$\Yssoutsepsystsigassocconttimecentvec$}
\newcommand{\Yssoutsepsystsigassocconttimecentvec}
           {\Yssoutsepsystsigconttimecentvec ^{\prime}}

\newcommand{\yssoutsepsystsigassocconttimecentvecval}
          {$\Yssoutsepsystsigassocconttimecentvecval$}
\newcommand{\Yssoutsepsystsigassocconttimecentvecval}
           {\Yssoutsepsystsigassocconttimecentvec (\Yssconttimeval)}

\newcommand{\yssoutsepsystsigassocconttimecentone}
          {$\Yssoutsepsystsigassocconttimecentone$}
\newcommand{\Yssoutsepsystsigassocconttimecentone}
           {\Yssoutsepsystsigassocconttimecentvec _{1}}

\newcommand{\yssoutsepsystsigassocconttimecentoneval}
          {$\Yssoutsepsystsigassocconttimecentoneval$}
\newcommand{\Yssoutsepsystsigassocconttimecentoneval}
           {\Yssoutsepsystsigassocconttimecentone (\Yssconttimeval)}

\newcommand{\yssoutsepsystsigassocconttimecentindex}
          {$\Yssoutsepsystsigassocconttimecentindex$}
\newcommand{\Yssoutsepsystsigassocconttimecentindex}
           {\Yssoutsepsystsigassocconttimecentvec _{\Ysswayindex}}

\newcommand{\yssoutsepsystsigassocconttimecentindexval}
          {$\Yssoutsepsystsigassocconttimecentindexval$}
\newcommand{\Yssoutsepsystsigassocconttimecentindexval}
           {\Yssoutsepsystsigassocconttimecentindex (\Yssconttimeval)}

\newcommand{\yssoutsepsystsigassocconttimecentindexother}
          {$\Yssoutsepsystsigassocconttimecentindexother$}
\newcommand{\Yssoutsepsystsigassocconttimecentindexother}
           {\Yssoutsepsystsigassocconttimecentvec _{\Ysswayindexother}}

\newcommand{\yssoutsepsystsigassocconttimecentindexotherval}
          {$\Yssoutsepsystsigassocconttimecentindexotherval$}
\newcommand{\Yssoutsepsystsigassocconttimecentindexotherval}
           {\Yssoutsepsystsigassocconttimecentindexother (\Yssconttimeval)}

\newcommand{\yssoutsepsystsigassocconttimecentindexfourth}
{$\Yssoutsepsystsigassocconttimecentindexfourth$}
\newcommand{\Yssoutsepsystsigassocconttimecentindexfourth}
{\Yssoutsepsystsigassocconttimecentvec _{\Ysswayindexfourth}}

\newcommand{\yssoutsepsystsigassocconttimecentindexfourthval}
{$\Yssoutsepsystsigassocconttimecentindexfourthval$}
\newcommand{\Yssoutsepsystsigassocconttimecentindexfourthval}
{\Yssoutsepsystsigassocconttimecentindexfourth (\Yssconttimeval)}


\newcommand{\yssoutsepsystsigdisctimecentvec}
          {$\Yssoutsepsystsigdisctimecentvec$}
\newcommand{\Yssoutsepsystsigdisctimecentvec}
           {y}

\newcommand{\yssoutsepsystsigdisctimecentvecval}
          {$\Yssoutsepsystsigdisctimecentvecval$}
\newcommand{\Yssoutsepsystsigdisctimecentvecval}
           {\Yssoutsepsystsigdisctimecentvec (\Yssdisctimeval)}

\newcommand{\yssoutsepsystsigdisctimecentvecvalother}
{$\Yssoutsepsystsigdisctimecentvecvalother$}
\newcommand{\Yssoutsepsystsigdisctimecentvecvalother}
{\Yssoutsepsystsigdisctimecentvec (\Yssdisctimevalother)}

\newcommand{\yssoutsepsystsigdisctimecentone}
          {$\Yssoutsepsystsigdisctimecentone$}
\newcommand{\Yssoutsepsystsigdisctimecentone}
           {\Yssoutsepsystsigdisctimecentvec _{1}}

\newcommand{\yssoutsepsystsigdisctimecentoneval}
          {$\Yssoutsepsystsigdisctimecentoneval$}
\newcommand{\Yssoutsepsystsigdisctimecentoneval}
           {\Yssoutsepsystsigdisctimecentone (\Yssdisctimeval)}

\newcommand{\yssoutsepsystsigdisctimecentonevalother}
{$\Yssoutsepsystsigdisctimecentonevalother$}
\newcommand{\Yssoutsepsystsigdisctimecentonevalother}
{\Yssoutsepsystsigdisctimecentone (\Yssdisctimevalother)}

\newcommand{\yssoutsepsystsigdisctimecenttwo}
{$\Yssoutsepsystsigdisctimecenttwo$}
\newcommand{\Yssoutsepsystsigdisctimecenttwo}
{\Yssoutsepsystsigdisctimecentvec _{2}}

\newcommand{\yssoutsepsystsigdisctimecenttwovalother}
{$\Yssoutsepsystsigdisctimecenttwovalother$}
\newcommand{\Yssoutsepsystsigdisctimecenttwovalother}
{\Yssoutsepsystsigdisctimecenttwo (\Yssdisctimevalother)}

\newcommand{\yssoutsepsystsigdisctimecentindex}
          {$\Yssoutsepsystsigdisctimecentindex$}
\newcommand{\Yssoutsepsystsigdisctimecentindex}
           {\Yssoutsepsystsigdisctimecentvec _{\Ysswayindex}}

\newcommand{\yssoutsepsystsigdisctimecentindexval}
          {$\Yssoutsepsystsigdisctimecentindexval$}
\newcommand{\Yssoutsepsystsigdisctimecentindexval}
           {\Yssoutsepsystsigdisctimecentindex (\Yssdisctimeval)}

\newcommand{\yssoutsepsystsigdisctimecentindexvalother}
{$\Yssoutsepsystsigdisctimecentindexvalother$}
\newcommand{\Yssoutsepsystsigdisctimecentindexvalother}
{\Yssoutsepsystsigdisctimecentindex (\Yssdisctimevalother)}

\newcommand{\yssoutsepsystsigdisctimecentindexother}
          {$\Yssoutsepsystsigdisctimecentindexother$}
\newcommand{\Yssoutsepsystsigdisctimecentindexother}
           {\Yssoutsepsystsigdisctimecentvec _{\Ysswayindexother}}

\newcommand{\yssoutsepsystsigdisctimecentindexotherval}
          {$\Yssoutsepsystsigdisctimecentindexotherval$}
\newcommand{\Yssoutsepsystsigdisctimecentindexotherval}
           {\Yssoutsepsystsigdisctimecentindexother (\Yssdisctimeval)}

\newcommand{\yssoutsepsystsigdisctimecentindexothervalother}
{$\Yssoutsepsystsigdisctimecentindexothervalother$}
\newcommand{\Yssoutsepsystsigdisctimecentindexothervalother}
{\Yssoutsepsystsigdisctimecentindexother (\Yssdisctimevalother)}

\newcommand{\yssoutsepsystsigdisctimecentindexsensnb}
          {$\Yssoutsepsystsigdisctimecentindexsensnb$}
\newcommand{\Yssoutsepsystsigdisctimecentindexsensnb}
           {\Yssoutsepsystsigdisctimecentvec _{\Ysssensnb}}

\newcommand{\yssoutsepsystsigdisctimecentindexsensnbval}
          {$\Yssoutsepsystsigdisctimecentindexsensnbval$}
\newcommand{\Yssoutsepsystsigdisctimecentindexsensnbval}
           {\Yssoutsepsystsigdisctimecentindexsensnb (\Yssdisctimeval)}


\newcommand{\yssoutsepsystsigassocdisctimecontfreqnotonenoncentvec}
          {$\Yssoutsepsystsigassocdisctimecontfreqnotonenoncentvec$}
\newcommand{\Yssoutsepsystsigassocdisctimecontfreqnotonenoncentvec}
           {Y
       ^{\prime}}

\newcommand{\yssoutsepsystsigassocdisctimecontfreqnotonenoncentvecval}
          {$\Yssoutsepsystsigassocdisctimecontfreqnotonenoncentvecval$}
\newcommand{\Yssoutsepsystsigassocdisctimecontfreqnotonenoncentvecval}
           {\Yssoutsepsystsigassocdisctimecontfreqnotonenoncentvec ( \Ysscontfreqval ) }


\newcommand{\yssoutsepsystsigdisctimecentveczt}
          {$\Yssoutsepsystsigdisctimecentveczt$}
\newcommand{\Yssoutsepsystsigdisctimecentveczt}
           {Y}

\newcommand{\yssoutsepsystsigdisctimecentvecztval}
          {$\Yssoutsepsystsigdisctimecentvecztval$}
\newcommand{\Yssoutsepsystsigdisctimecentvecztval}
           {\Yssoutsepsystsigdisctimecentveczt (z)}

\newcommand{\yssoutsepsystsigdisctimecentonezt}
          {$\Yssoutsepsystsigdisctimecentonezt$}
\newcommand{\Yssoutsepsystsigdisctimecentonezt}
           {\Yssoutsepsystsigdisctimecentveczt _{1}}

\newcommand{\yssoutsepsystsigdisctimecentoneztval}
          {$\Yssoutsepsystsigdisctimecentoneztval$}
\newcommand{\Yssoutsepsystsigdisctimecentoneztval}
           {\Yssoutsepsystsigdisctimecentonezt (z)}

\newcommand{\yssoutsepsystsigdisctimecenttwozt}
          {$\Yssoutsepsystsigdisctimecenttwozt$}
\newcommand{\Yssoutsepsystsigdisctimecenttwozt}
           {\Yssoutsepsystsigdisctimecentveczt _{2}}

\newcommand{\yssoutsepsystsigdisctimecenttwoztval}
          {$\Yssoutsepsystsigdisctimecenttwoztval$}
\newcommand{\Yssoutsepsystsigdisctimecenttwoztval}
           {\Yssoutsepsystsigdisctimecenttwozt (z)}

\newcommand{\yssoutsepsystsigdisctimecentindexzt}
          {$\Yssoutsepsystsigdisctimecentindexzt$}
\newcommand{\Yssoutsepsystsigdisctimecentindexzt}
           {\Yssoutsepsystsigdisctimecentveczt _{\Ysswayindex}}

\newcommand{\yssoutsepsystsigdisctimecentindexztval}
          {$\Yssoutsepsystsigdisctimecentindexztval$}
\newcommand{\Yssoutsepsystsigdisctimecentindexztval}
           {\Yssoutsepsystsigdisctimecentindexzt (z)}

\newcommand{\yssoutsepsystsigdisctimecentindexsensnbzt}
          {$\Yssoutsepsystsigdisctimecentindexsensnbzt$}
\newcommand{\Yssoutsepsystsigdisctimecentindexsensnbzt}
           {\Yssoutsepsystsigdisctimecentveczt _{\Ysssensnb}}

\newcommand{\yssoutsepsystsigdisctimecentindexsensnbztval}
          {$\Yssoutsepsystsigdisctimecentindexsensnbztval$}
\newcommand{\Yssoutsepsystsigdisctimecentindexsensnbztval}
           {\Yssoutsepsystsigdisctimecentindexsensnbzt (z)}



\newcommand{\yssarbsigconttimenotone}
          {$\Yssarbsigconttimenotone$}
\newcommand{\Yssarbsigconttimenotone}
           {v}

\newcommand{\yssarbsigconttimenotoneval}
          {$\Yssarbsigconttimenotoneval$}
\newcommand{\Yssarbsigconttimenotoneval}
           {\Yssarbsigconttimenotone (\Yssconttimeval)}

\newcommand{\yssarbsigconttimenotoneindexone}
          {$\Yssarbsigconttimenotoneindexone$}
\newcommand{\Yssarbsigconttimenotoneindexone}
           {\Yssarbsigconttimenotone _{1}}

\newcommand{\yssarbsigconttimenotoneindexoneval}
          {$\Yssarbsigconttimenotoneindexoneval$}
\newcommand{\Yssarbsigconttimenotoneindexoneval}
           {\Yssarbsigconttimenotoneindexone (\Yssconttimeval)}

\newcommand{\yssarbsigconttimenotoneindextwo}
          {$\Yssarbsigconttimenotoneindextwo$}
\newcommand{\Yssarbsigconttimenotoneindextwo}
           {\Yssarbsigconttimenotone _{2}}

\newcommand{\yssarbsigconttimenotoneindextwoval}
          {$\Yssarbsigconttimenotoneindextwoval$}
\newcommand{\Yssarbsigconttimenotoneindextwoval}
           {\Yssarbsigconttimenotoneindextwo (\Yssconttimeval)}

\newcommand{\yssarbsigconttimenottwo}
          {$\Yssarbsigconttimenottwo$}
\newcommand{\Yssarbsigconttimenottwo}
           {w}


\newcommand{\yssarbsigconttimecontfreqnoncentnotone}
          {$\Yssarbsigconttimecontfreqnoncentnotone$}
\newcommand{\Yssarbsigconttimecontfreqnoncentnotone}
           {V}

\newcommand{\yssarbsigconttimecontfreqnoncentnotoneval}
          {$\Yssarbsigconttimecontfreqnoncentnotoneval$}
\newcommand{\Yssarbsigconttimecontfreqnoncentnotoneval}
           {\Yssarbsigconttimecontfreqnoncentnotone (\Yssconttimeval , \Ysscontfreqval)}

\newcommand{\yssarbsigconttimecontfreqnoncentnotoneindexone}
          {$\Yssarbsigconttimecontfreqnoncentnotoneindexone$}
\newcommand{\Yssarbsigconttimecontfreqnoncentnotoneindexone}
           {\Yssarbsigconttimecontfreqnoncentnotone _{1}}

\newcommand{\yssarbsigconttimecontfreqnoncentnotonevalindexone}
          {$\Yssarbsigconttimecontfreqnoncentnotonevalindexone$}
\newcommand{\Yssarbsigconttimecontfreqnoncentnotonevalindexone}
           {\Yssarbsigconttimecontfreqnoncentnotoneindexone (\Yssconttimeval , \Ysscontfreqval)}

\newcommand{\yssarbsigconttimecontfreqnoncentnotoneindextwo}
          {$\Yssarbsigconttimecontfreqnoncentnotoneindextwo$}
\newcommand{\Yssarbsigconttimecontfreqnoncentnotoneindextwo}
           {\Yssarbsigconttimecontfreqnoncentnotone _{2}}

\newcommand{\yssarbsigconttimecontfreqnoncentnotonevalindextwo}
          {$\Yssarbsigconttimecontfreqnoncentnotonevalindextwo$}
\newcommand{\Yssarbsigconttimecontfreqnoncentnotonevalindextwo}
           {\Yssarbsigconttimecontfreqnoncentnotoneindextwo (\Yssconttimeval , \Ysscontfreqval)}

\newcommand{\yssarbsigconttimecontfreqnoncentnottwo}
          {$\Yssarbsigconttimecontfreqnoncentnottwo$}
\newcommand{\Yssarbsigconttimecontfreqnoncentnottwo}
           {W}


\newcommand{\yssarbsigconttimecontscalenoncentnotone}
          {$\Yssarbsigconttimecontscalenoncentnotone$}
\newcommand{\Yssarbsigconttimecontscalenoncentnotone}
           {\Ysstimescalecontcoefnot _{\Yssarbsigconttimenotone}}

\newcommand{\yssarbsigconttimecontscalenoncentnotoneval}
          {$\Yssarbsigconttimecontscalenoncentnotoneval$}
\newcommand{\Yssarbsigconttimecontscalenoncentnotoneval}
           {\Yssarbsigconttimecontscalenoncentnotone
        ( \Ysstimescalecontshiftval , \Ysstimescalecontscaleval )
           }

\newcommand{\yssarbsigconttimecontscalenoncentnotoneindexone}
          {$\Yssarbsigconttimecontscalenoncentnotoneindexone$}
\newcommand{\Yssarbsigconttimecontscalenoncentnotoneindexone}
           {\Ysstimescalecontcoefnot _{\Yssarbsigconttimenotoneindexone} }

\newcommand{\yssarbsigconttimecontscalenoncentnotonevalindexone}
          {$\Yssarbsigconttimecontscalenoncentnotonevalindexone$}
\newcommand{\Yssarbsigconttimecontscalenoncentnotonevalindexone}
           {\Yssarbsigconttimecontscalenoncentnotoneindexone
        ( \Ysstimescalecontshiftval , \Ysstimescalecontscaleval )
       }

\newcommand{\yssarbsigconttimecontscalenoncentnotoneindextwo}
          {$\Yssarbsigconttimecontscalenoncentnotoneindextwo$}
\newcommand{\Yssarbsigconttimecontscalenoncentnotoneindextwo}
           {\Ysstimescalecontcoefnot _{\Yssarbsigconttimenotoneindextwo} }

\newcommand{\yssarbsigconttimecontscalenoncentnotonevalindextwo}
          {$\Yssarbsigconttimecontscalenoncentnotonevalindextwo$}
\newcommand{\Yssarbsigconttimecontscalenoncentnotonevalindextwo}
           {\Yssarbsigconttimecontscalenoncentnotoneindextwo
        ( \Ysstimescalecontshiftval , \Ysstimescalecontscaleval )
       }


\newcommand{\yssarbsigdisctimenotone}
          {$\Yssarbsigdisctimenotone$}
\newcommand{\Yssarbsigdisctimenotone}
           {v}

\newcommand{\yssarbsigdisctimenotoneval}
          {$\Yssarbsigdisctimenotoneval$}
\newcommand{\Yssarbsigdisctimenotoneval}
           {\Yssarbsigdisctimenotone (\Yssdisctimeval)}

\newcommand{\yssarbsigdisctimenottwo}
          {$\Yssarbsigdisctimenottwo$}
\newcommand{\Yssarbsigdisctimenottwo}
           {w}



\newcommand{\yssinnovtosrcfillengthnegnoindex}
           {$\Yssinnovtosrcfillengthnegnoindex$}
\newcommand{\Yssinnovtosrcfillengthnegnoindex}
           {L_{1}}

\newcommand{\yssinnovtosrcfillengthposnoindex}
           {$\Yssinnovtosrcfillengthposnoindex$}
\newcommand{\Yssinnovtosrcfillengthposnoindex}
           {L_{2}}


\newcommand{\yssinnovtosrcimprespnoindex}
           {$\Yssinnovtosrcimprespnoindex$}
\newcommand{\Yssinnovtosrcimprespnoindex}
           {d}

\newcommand{\yssinnovtosrcimprespone}
           {$\Yssinnovtosrcimprespone$}
\newcommand{\Yssinnovtosrcimprespone}
           {\Yssinnovtosrcimprespnoindex _{1}}

\newcommand{\yssinnovtosrcimpresptwo}
           {$\Yssinnovtosrcimpresptwo$}
\newcommand{\Yssinnovtosrcimpresptwo}
           {\Yssinnovtosrcimprespnoindex_{2}}

\newcommand{\yssinnovtosrcimprespindex}
           {$\Yssinnovtosrcimprespindex$}
\newcommand{\Yssinnovtosrcimprespindex}
           {\Yssinnovtosrcimprespnoindex_{\Ysswayindex}}


\newcommand{\yssinnovtosrctransfuncnoindex}
           {$\Yssinnovtosrctransfuncnoindex$}
\newcommand{\Yssinnovtosrctransfuncnoindex}
           {D}

\newcommand{\yssinnovtosrctransfuncindex}
           {$\Yssinnovtosrctransfuncindex$}
\newcommand{\Yssinnovtosrctransfuncindex}
           {\Yssinnovtosrctransfuncnoindex_{\Ysswayindex}}

\newcommand{\yssinnovtosrctransfuncindexval}
           {$\Yssinnovtosrctransfuncindexval$}
\newcommand{\Yssinnovtosrctransfuncindexval}
           {\Yssinnovtosrctransfuncindex (z)}



\newcommand{\yssmixmatrixscalar}
           {$\Yssmixmatrixscalar$}
\newcommand{\Yssmixmatrixscalar}
           {A}

\newcommand{\yssmixmatrixscalarestim}
           {$\Yssmixmatrixscalarestim$}
\newcommand{\Yssmixmatrixscalarestim}
           {\hat{\Yssmixmatrixscalar}}

\newcommand{\yssmixmatrixscalarelnoindex}
          {$\Yssmixmatrixscalarelnoindex$}
\newcommand{\Yssmixmatrixscalarelnoindex}
           {a}

\newcommand{\yssmixmatrixscalareloneone}
          {$\Yssmixmatrixscalareloneone$}
\newcommand{\Yssmixmatrixscalareloneone}
           {\Yssmixmatrixscalarelnoindex _{1 1}}

\newcommand{\yssmixmatrixscalarelonetwo}
          {$\Yssmixmatrixscalarelonetwo$}
\newcommand{\Yssmixmatrixscalarelonetwo}
           {\Yssmixmatrixscalarelnoindex _{1 2}}

\newcommand{\yssmixmatrixscalareloneindexother}
          {$\Yssmixmatrixscalareloneindexother$}
\newcommand{\Yssmixmatrixscalareloneindexother}
           {\Yssmixmatrixscalarelnoindex _{1 \Ysswayindexother}}

\newcommand{\yssmixmatrixscalareloneindexthird}
          {$\Yssmixmatrixscalareloneindexthird$}
\newcommand{\Yssmixmatrixscalareloneindexthird}
           {\Yssmixmatrixscalarelnoindex _{1 \Ysswayindexthird}}

\newcommand{\yssmixmatrixscalareloneindexfourth}
{$\Yssmixmatrixscalareloneindexfourth$}
\newcommand{\Yssmixmatrixscalareloneindexfourth}
{\Yssmixmatrixscalarelnoindex _{1 \Ysswayindexfourth}}

\newcommand{\yssmixmatrixscalareltwoone}
          {$\Yssmixmatrixscalareltwoone$}
\newcommand{\Yssmixmatrixscalareltwoone}
           {\Yssmixmatrixscalarelnoindex _{2 1}}

\newcommand{\yssmixmatrixscalareltwotwo}
          {$\Yssmixmatrixscalareltwotwo$}
\newcommand{\Yssmixmatrixscalareltwotwo}
           {\Yssmixmatrixscalarelnoindex _{2 2}}

\newcommand{\yssmixmatrixscalarelindexone}
          {$\Yssmixmatrixscalarelindexone$}
\newcommand{\Yssmixmatrixscalarelindexone}
           {\Yssmixmatrixscalarelnoindex _{\Ysswayindex 1}}

\newcommand{\yssmixmatrixscalarelindextwo}
          {$\Yssmixmatrixscalarelindextwo$}
\newcommand{\Yssmixmatrixscalarelindextwo}
           {\Yssmixmatrixscalarelnoindex _{\Ysswayindex 2}}

\newcommand{\yssmixmatrixscalarelindexthree}
          {$\Yssmixmatrixscalarelindexthree$}
\newcommand{\Yssmixmatrixscalarelindexthree}
           {\Yssmixmatrixscalarelnoindex _{\Ysswayindex 3}}

\newcommand{\yssmixmatrixscalarelindexindex}
          {$\Yssmixmatrixscalarelindexindex$}
\newcommand{\Yssmixmatrixscalarelindexindex}
           {\Yssmixmatrixscalarelnoindex _{\Ysswayindex \Ysswayindex}}

\newcommand{\yssmixmatrixscalarelindexindexother}
          {$\Yssmixmatrixscalarelindexindexother$}
\newcommand{\Yssmixmatrixscalarelindexindexother}
           {\Yssmixmatrixscalarelnoindex _{\Ysswayindex \Ysswayindexother}}

\newcommand{\yssmixmatrixscalarelindexindexthird}
          {$\Yssmixmatrixscalarelindexindexthird$}
\newcommand{\Yssmixmatrixscalarelindexindexthird}
           {\Yssmixmatrixscalarelnoindex _{\Ysswayindex \Ysswayindexthird}}

\newcommand{\yssmixmatrixscalarelindexindexfourth}
{$\Yssmixmatrixscalarelindexindexfourth$}
\newcommand{\Yssmixmatrixscalarelindexindexfourth}
{\Yssmixmatrixscalarelnoindex _{\Ysswayindex \Ysswayindexfourth}}

\newcommand{\yssmixmatrixscalarelindexotherindexother}
{$\Yssmixmatrixscalarelindexotherindexother$}
\newcommand{\Yssmixmatrixscalarelindexotherindexother}
{\Yssmixmatrixscalarelnoindex _{\Ysswayindexother \Ysswayindexother}}

\newcommand{\yssmixmatrixscalarelindexthirdindexother}
          {$\Yssmixmatrixscalarelindexthirdindexother$}
\newcommand{\Yssmixmatrixscalarelindexthirdindexother}
           {\Yssmixmatrixscalarelnoindex _{\Ysswayindexthird \Ysswayindexother}}

\newcommand{\yssmixmatrixscalarelindexthirdindexfourth}
          {$\Yssmixmatrixscalarelindexthirdindexfourth$}
\newcommand{\Yssmixmatrixscalarelindexthirdindexfourth}
           {\Yssmixmatrixscalarelnoindex _{\Ysswayindexthird \Ysswayindexfourth}}

\newcommand{\yssinvmixmatrixscalar}
           {$\Yssinvmixmatrixscalar$}
\newcommand{\Yssinvmixmatrixscalar}
           {\Yssmixmatrixscalar ^{-1}}

\newcommand{\yssinvmixmatrixscalarestim}
           {$\Yssinvmixmatrixscalarestim$}
\newcommand{\Yssinvmixmatrixscalarestim}
           {\Yssmixmatrixscalarestim ^{-1}}



\newcommand{\yssmixmatrixlagellagnoindex}
          {$\Yssmixmatrixlagellagnoindex$}
\newcommand{\Yssmixmatrixlagellagnoindex}
           {n}

\newcommand{\yssmixmatrixlagellagindexindexother}
          {$\Yssmixmatrixlagellagindexindexother$}
\newcommand{\Yssmixmatrixlagellagindexindexother}
           {\Yssmixmatrixlagellagnoindex _{\Ysswayindex \Ysswayindexother}}

\newcommand{\yssmixmatrixlagellagoneindexthird}
          {$\Yssmixmatrixlagellagoneindexthird$}
\newcommand{\Yssmixmatrixlagellagoneindexthird}
           {\Yssmixmatrixlagellagnoindex _{1 \Ysswayindexthird}}

\newcommand{\yssmixmatrixlagellagindexindexthird}
          {$\Yssmixmatrixlagellagindexindexthird$}
\newcommand{\Yssmixmatrixlagellagindexindexthird}
           {\Yssmixmatrixlagellagnoindex _{\Ysswayindex \Ysswayindexthird}}



\newcommand{\yssmixmatrixcontfreqnotonefunc}
           {$\Yssmixmatrixcontfreqnotonefunc$}
\newcommand{\Yssmixmatrixcontfreqnotonefunc}
           {A}


\newcommand{\yssmixmatrixcontfreqnotoneval}
           {$\Yssmixmatrixcontfreqnotoneval$}
\newcommand{\Yssmixmatrixcontfreqnotoneval}
           {\Yssmixmatrixcontfreqnotonefunc ( \Ysscontfreqval ) }




\newcommand{\yssmixtransfuncorder}
          {$\Yssmixtransfuncorder$}
\newcommand{\Yssmixtransfuncorder}
           {M}


\newcommand{\ymixmatrixzt}
           {$\Ymixmatrixzt$}
\newcommand{\Ymixmatrixzt}
           {A(z)}


\newcommand{\yssmixtransfuncnoindex}
           {$\Yssmixtransfuncnoindex$}
\newcommand{\Yssmixtransfuncnoindex}
           {A}

\newcommand{\yssmixtransfunconeone}
          {$\Yssmixtransfunconeone$}
\newcommand{\Yssmixtransfunconeone}
           {\Yssmixtransfuncnoindex _{1 1} (z)}

\newcommand{\yssmixtransfunconetwo}
          {$\Yssmixtransfunconetwo$}
\newcommand{\Yssmixtransfunconetwo}
           {\Yssmixtransfuncnoindex _{1 2} (z)}

\newcommand{\yssmixtransfunconethree}
          {$\Yssmixtransfunconethree$}
\newcommand{\Yssmixtransfunconethree}
           {\Yssmixtransfuncnoindex _{1 3} (z)}

\newcommand{\yssmixtransfunctwoone}
          {$\Yssmixtransfunctwoone$}
\newcommand{\Yssmixtransfunctwoone}
           {\Yssmixtransfuncnoindex _{2 1} (z)}

\newcommand{\yssmixtransfunctwotwo}
          {$\Yssmixtransfunctwotwo$}
\newcommand{\Yssmixtransfunctwotwo}
           {\Yssmixtransfuncnoindex _{2 2} (z)}

\newcommand{\yssmixtransfunctwothree}
          {$\Yssmixtransfunctwothree$}
\newcommand{\Yssmixtransfunctwothree}
           {\Yssmixtransfuncnoindex _{2 3} (z)}

\newcommand{\yssmixtransfunconeindexother}
          {$\Yssmixtransfunconeindexother$}
\newcommand{\Yssmixtransfunconeindexother}
           {\Yssmixtransfuncnoindex _{1 \Ysswayindexother} (z)}

\newcommand{\yssmixtransfunctwoindexother}
          {$\Yssmixtransfunctwoindexother$}
\newcommand{\Yssmixtransfunctwoindexother}
           {\Yssmixtransfuncnoindex _{2 \Ysswayindexother} (z)}

\newcommand{\yssmixtransfuncindexindexother}
          {$\Yssmixtransfuncindexindexother$}
\newcommand{\Yssmixtransfuncindexindexother}
           {\Yssmixtransfuncnoindex _{\Ysswayindex \Ysswayindexother} (z)}


\newcommand{\yssmixmatrixscalarquad}
{$\Yssmixmatrixscalarquad$}
\newcommand{\Yssmixmatrixscalarquad}
{Q}

\newcommand{\yssmixmatrixscalarelquadnoindex}
          {$\Yssmixmatrixscalarelquadnoindex$}
\newcommand{\Yssmixmatrixscalarelquadnoindex}
           {q}

\newcommand{\yssmixmatrixscalarelquadone}
{$\Yssmixmatrixscalarelquadone$}
\newcommand{\Yssmixmatrixscalarelquadone}
{\Yssmixmatrixscalarelquadnoindex _{1}}

\newcommand{\yssmixmatrixscalarelquadtwo}
{$\Yssmixmatrixscalarelquadtwo$}
\newcommand{\Yssmixmatrixscalarelquadtwo}
{\Yssmixmatrixscalarelquadnoindex _{2}}

\newcommand{\yssmixmatrixscalarelquadindex}
{$\Yssmixmatrixscalarelquadindex$}
\newcommand{\Yssmixmatrixscalarelquadindex}
{\Yssmixmatrixscalarelquadnoindex _{\Ysswayindex}}

\newcommand{\yssmixmatrixscalarelquadoneindexotherindexthird}
{$\Yssmixmatrixscalarelquadoneindexotherindexthird$}
\newcommand{\Yssmixmatrixscalarelquadoneindexotherindexthird}
{\Yssmixmatrixscalarelquadnoindex _{1 \Ysswayindexother \Ysswayindexthird}}

\newcommand{\yssmixmatrixscalarelquadindexindexotherindexthird}
{$\Yssmixmatrixscalarelquadindexindexotherindexthird$}
\newcommand{\Yssmixmatrixscalarelquadindexindexotherindexthird}
{\Yssmixmatrixscalarelquadnoindex _{\Ysswayindex \Ysswayindexother \Ysswayindexthird}}


\newcommand{\yssmixmatrixscalarlinquadpartlinelnoindex}
{$\Yssmixmatrixscalarlinquadpartlinelnoindex$}
\newcommand{\Yssmixmatrixscalarlinquadpartlinelnoindex}
{L}

\newcommand{\yssmixmatrixscalarlinquadpartlinelonetwo}
{$\Yssmixmatrixscalarlinquadpartlinelonetwo$}
\newcommand{\Yssmixmatrixscalarlinquadpartlinelonetwo}
{\Yssmixmatrixscalarlinquadpartlinelnoindex_{12}}

\newcommand{\yssmixmatrixscalarlinquadpartlineltwoone}
{$\Yssmixmatrixscalarlinquadpartlineltwoone$}
\newcommand{\Yssmixmatrixscalarlinquadpartlineltwoone}
{\Yssmixmatrixscalarlinquadpartlinelnoindex_{21}}

\newcommand{\yssmixmatrixscalarlinquadpartlinelindexindexother}
{$\Yssmixmatrixscalarlinquadpartlinelindexindexother$}
\newcommand{\Yssmixmatrixscalarlinquadpartlinelindexindexother}
{\Yssmixmatrixscalarlinquadpartlinelnoindex_{\Ysswayindex \Ysswayindexother}}

\newcommand{\yssmixmatrixscalarlinquadpartquadelnoindex}
{$\Yssmixmatrixscalarlinquadpartquadelnoindex$}
\newcommand{\Yssmixmatrixscalarlinquadpartquadelnoindex}
{Q}

\newcommand{\yssmixmatrixscalarlinquadpartquadelone}
{$\Yssmixmatrixscalarlinquadpartquadelone$}
\newcommand{\Yssmixmatrixscalarlinquadpartquadelone}
{\Yssmixmatrixscalarlinquadpartquadelnoindex_{1}}

\newcommand{\yssmixmatrixscalarlinquadpartquadeltwo}
{$\Yssmixmatrixscalarlinquadpartquadeltwo$}
\newcommand{\Yssmixmatrixscalarlinquadpartquadeltwo}
{\Yssmixmatrixscalarlinquadpartquadelnoindex_{2}}

\newcommand{\yssmixmatrixscalarlinquadpartquadelindex}
{$\Yssmixmatrixscalarlinquadpartquadelindex$}
\newcommand{\Yssmixmatrixscalarlinquadpartquadelindex}
{\Yssmixmatrixscalarlinquadpartquadelnoindex_{\Ysswayindex}}



\newcommand{\ysssepsystmatrixscalar}
          {$\Ysssepsystmatrixscalar$}
\newcommand{\Ysssepsystmatrixscalar}
           {B}

\newcommand{\ysssepsystmatrixscalarelnoindex}
          {$\Ysssepsystmatrixscalarelnoindex$}
\newcommand{\Ysssepsystmatrixscalarelnoindex}
{w}

\newcommand{\ysssepsystmatrixscalareloneindexother}
          {$\Ysssepsystmatrixscalareloneindexother$}
\newcommand{\Ysssepsystmatrixscalareloneindexother}
           {\Ysssepsystmatrixscalarelnoindex _{1 \Ysswayindexother}}

\newcommand{\ysssepsystmatrixscalarelindexindexother}
          {$\Ysssepsystmatrixscalarelindexindexother$}
\newcommand{\Ysssepsystmatrixscalarelindexindexother}
           {\Ysssepsystmatrixscalarelnoindex _{\Ysswayindex \Ysswayindexother}}


\newcommand{\ysssepsystmatrixscalarelonetwo}
          {$\Ysssepsystmatrixscalarelonetwo$}
\newcommand{\Ysssepsystmatrixscalarelonetwo}
           {\Ysssepsystmatrixscalarelnoindex _{1 2}}

\newcommand{\ysssepsystmatrixscalareltwoone}
          {$\Ysssepsystmatrixscalareltwoone$}
\newcommand{\Ysssepsystmatrixscalareltwoone}
           {\Ysssepsystmatrixscalarelnoindex _{2 1}}



\newcommand{\ysssepsysmatrixscalarassocone}
           {$\Ysssepsysmatrixscalarassocone$}
\newcommand{\Ysssepsysmatrixscalarassocone}
           {C}

\newcommand{\ysssepsystmatrixscalarassoconeelnoindex}
          {$\Ysssepsystmatrixscalarassoconeelnoindex$}
\newcommand{\Ysssepsystmatrixscalarassoconeelnoindex}
           {\alpha}

\newcommand{\ysssepsystmatrixscalarassoconeeltwoone}
          {$\Ysssepsystmatrixscalarassoconeeltwoone$}
\newcommand{\Ysssepsystmatrixscalarassoconeeltwoone}
           {\Ysssepsystmatrixscalarassoconeelnoindex _{2 1}}

\newcommand{\ysssepsystmatrixscalarassoconeeltwotwo}
          {$\Ysssepsystmatrixscalarassoconeeltwotwo$}
\newcommand{\Ysssepsystmatrixscalarassoconeeltwotwo}
           {\Ysssepsystmatrixscalarassoconeelnoindex _{2 2}}

\newcommand{\ysssepsystmatrixscalarassoconeelindexindexother}
          {$\Ysssepsystmatrixscalarassoconeelindexindexother$}
\newcommand{\Ysssepsystmatrixscalarassoconeelindexindexother}
           {\Ysssepsystmatrixscalarassoconeelnoindex _{\Ysswayindex \Ysswayindexother}}

\newcommand{\ysssepsystmatrixscalarassoconeelindexindexthird}
          {$\Ysssepsystmatrixscalarassoconeelindexindexthird$}
\newcommand{\Ysssepsystmatrixscalarassoconeelindexindexthird}
           {\Ysssepsystmatrixscalarassoconeelnoindex _{\Ysswayindex \Ysswayindexthird}}


\newcommand{\ysssepsysmatrixscalarassoctwo}
           {$\Ysssepsysmatrixscalarassoctwo$}
\newcommand{\Ysssepsysmatrixscalarassoctwo}
           {D}

\newcommand{\ysssepsystmatrixscalarassoctwoelnoindex}
          {$\Ysssepsystmatrixscalarassoctwoelnoindex$}
\newcommand{\Ysssepsystmatrixscalarassoctwoelnoindex}
           {\beta}

\newcommand{\ysssepsystmatrixscalarassoctwoeltwoone}
          {$\Ysssepsystmatrixscalarassoctwoeltwoone$}
\newcommand{\Ysssepsystmatrixscalarassoctwoeltwoone}
           {\Ysssepsystmatrixscalarassoctwoelnoindex _{2 1}}

\newcommand{\ysssepsystmatrixscalarassoctwoeltwotwo}
          {$\Ysssepsystmatrixscalarassoctwoeltwotwo$}
\newcommand{\Ysssepsystmatrixscalarassoctwoeltwotwo}
           {\Ysssepsystmatrixscalarassoctwoelnoindex _{2 2}}

\newcommand{\ysssepsystmatrixscalarassoctwoelindexindexother}
          {$\Ysssepsystmatrixscalarassoctwoelindexindexother$}
\newcommand{\Ysssepsystmatrixscalarassoctwoelindexindexother}
           {\Ysssepsystmatrixscalarassoctwoelnoindex _{\Ysswayindex \Ysswayindexother}}

\newcommand{\ysssepsystmatrixscalarassoctwoelindexindexthird}
          {$\Ysssepsystmatrixscalarassoctwoelindexindexthird$}
\newcommand{\Ysssepsystmatrixscalarassoctwoelindexindexthird}
           {\Ysssepsystmatrixscalarassoctwoelnoindex _{\Ysswayindex \Ysswayindexthird}}


\newcommand{\ysssepsysmatrixscalarassocthree}
           {$\Ysssepsysmatrixscalarassocthree$}
\newcommand{\Ysssepsysmatrixscalarassocthree}
           {\Ysssepsysmatrixscalarassocone^{\prime}}

\newcommand{\ysssepsystmatrixscalarassocthreeelnoindex}
          {$\Ysssepsystmatrixscalarassocthreeelnoindex$}
\newcommand{\Ysssepsystmatrixscalarassocthreeelnoindex}
           {\Ysssepsystmatrixscalarassoconeelnoindex^{\prime}}

\newcommand{\ysssepsystmatrixscalarassocthreeelindexindexother}
          {$\Ysssepsystmatrixscalarassocthreeelindexindexother$}
\newcommand{\Ysssepsystmatrixscalarassocthreeelindexindexother}
           {\Ysssepsystmatrixscalarassocthreeelnoindex _{\Ysswayindex \Ysswayindexother}}

\newcommand{\ysssepsystmatrixscalarassocthreeelindexindexthird}
          {$\Ysssepsystmatrixscalarassocthreeelindexindexthird$}
\newcommand{\Ysssepsystmatrixscalarassocthreeelindexindexthird}
           {\Ysssepsystmatrixscalarassocthreeelnoindex _{\Ysswayindex \Ysswayindexthird}}


\newcommand{\ysssepsystmatrixcontfreqnotonefunc}
           {$\Ysssepsystmatrixcontfreqnotonefunc$}
\newcommand{\Ysssepsystmatrixcontfreqnotonefunc}
           {B}

\newcommand{\ysssepsystmatrixcontfreqnotoneval}
           {$\Ysssepsystmatrixcontfreqnotoneval$}
\newcommand{\Ysssepsystmatrixcontfreqnotoneval}
           {\Ysssepsystmatrixcontfreqnotonefunc ( \Ysscontfreqval ) }

\newcommand{\ysssepsystindexindexothercontfreqnotoneval}
          {$\Ysssepsystindexindexothercontfreqnotoneval$}
\newcommand{\Ysssepsystindexindexothercontfreqnotoneval}
           {\Ysssepsystmatrixcontfreqnotonefunc
        _{\Ysswayindex \Ysswayindexother} ( \Ysscontfreqval )}




\newcommand{\ysssepsysttransfuncnoindex}
           {$\Ysssepsysttransfuncnoindex$}
\newcommand{\Ysssepsysttransfuncnoindex}
           {B}

\newcommand{\ysssepsysttransfuncnoindexval}
           {$\Ysssepsysttransfuncnoindexval$}
\newcommand{\Ysssepsysttransfuncnoindexval}
           {\Ysssepsysttransfuncnoindex (z)}

\newcommand{\ysssepsysttransfunconetwo}
          {$\Ysssepsysttransfunconetwo$}
\newcommand{\Ysssepsysttransfunconetwo}
           {\Ysssepsysttransfuncnoindex _{1 2} (z)}

\newcommand{\ysssepsysttransfunctwoone}
          {$\Ysssepsysttransfunctwoone$}
\newcommand{\Ysssepsysttransfunctwoone}
           {\Ysssepsysttransfuncnoindex _{2 1} (z)}

\newcommand{\ysssepsysttransfuncindexindexother}
          {$\Ysssepsysttransfuncindexindexother$}
\newcommand{\Ysssepsysttransfuncindexindexother}
           {\Ysssepsysttransfuncnoindex _{\Ysswayindex \Ysswayindexother} (z)}


\newcommand{\ysssepsystcoefnoindex}
           {$\Ysssepsystcoefnoindex$}
\newcommand{\Ysssepsystcoefnoindex}
           {b}

\newcommand{\ysssepsystcoefindexindexother}
           {$\Ysssepsystcoefindexindexother$}
\newcommand{\Ysssepsystcoefindexindexother}
           {\Ysssepsystcoefnoindex _{\Ysswayindex \Ysswayindexother}}

\newcommand{\ysssepsystcoefonetwo}
           {$\Ysssepsystcoefonetwo$}
\newcommand{\Ysssepsystcoefonetwo}
           {\Ysssepsystcoefnoindex _{1 2}}

\newcommand{\ysssepsystcoeftwoone}
           {$\Ysssepsystcoeftwoone$}
\newcommand{\Ysssepsystcoeftwoone}
           {\Ysssepsystcoefnoindex _{2 1}}


\newcommand{\ysssepsystscalarlinquadpartlinelnoindex}
{$\Ysssepsystscalarlinquadpartlinelnoindex$}
\newcommand{\Ysssepsystscalarlinquadpartlinelnoindex}
{l}

\newcommand{\ysssepsystscalarlinquadpartlineloneone}
{$\Ysssepsystscalarlinquadpartlineloneone$}
\newcommand{\Ysssepsystscalarlinquadpartlineloneone}
{\Ysssepsystscalarlinquadpartlinelnoindex_{11}}

\newcommand{\ysssepsystscalarlinquadpartlineloneoneassoc}
{$\Ysssepsystscalarlinquadpartlineloneoneassoc$}
\newcommand{\Ysssepsystscalarlinquadpartlineloneoneassoc}
{\Ysssepsystscalarlinquadpartlineloneone ^{\prime}}

\newcommand{\ysssepsystscalarlinquadpartlinelonetwo}
{$\Ysssepsystscalarlinquadpartlinelonetwo$}
\newcommand{\Ysssepsystscalarlinquadpartlinelonetwo}
{\Ysssepsystscalarlinquadpartlinelnoindex_{12}}

\newcommand{\ysssepsystscalarlinquadpartlineltwoone}
{$\Ysssepsystscalarlinquadpartlineltwoone$}
\newcommand{\Ysssepsystscalarlinquadpartlineltwoone}
{\Ysssepsystscalarlinquadpartlinelnoindex_{21}}

\newcommand{\ysssepsystscalarlinquadpartlineltwotwo}
{$\Ysssepsystscalarlinquadpartlineltwotwo$}
\newcommand{\Ysssepsystscalarlinquadpartlineltwotwo}
{\Ysssepsystscalarlinquadpartlinelnoindex_{22}}

\newcommand{\ysssepsystscalarlinquadpartlineltwotwoassoc}
{$\Ysssepsystscalarlinquadpartlineltwotwoassoc$}
\newcommand{\Ysssepsystscalarlinquadpartlineltwotwoassoc}
{\Ysssepsystscalarlinquadpartlineltwotwo ^{\prime}}

\newcommand{\ysssepsystscalarlinquadpartlinelindexindex}
{$\Ysssepsystscalarlinquadpartlinelindexindex$}
\newcommand{\Ysssepsystscalarlinquadpartlinelindexindex}
{\Ysssepsystscalarlinquadpartlinelnoindex_{\Ysswayindex \Ysswayindex}}

\newcommand{\ysssepsystscalarlinquadpartlinelindexindexassoc}
{$\Ysssepsystscalarlinquadpartlinelindexindexassoc$}
\newcommand{\Ysssepsystscalarlinquadpartlinelindexindexassoc}
{\Ysssepsystscalarlinquadpartlinelindexindex ^{\prime}}

\newcommand{\ysssepsystscalarlinquadpartlinelindexindexother}
{$\Ysssepsystscalarlinquadpartlinelindexindexother$}
\newcommand{\Ysssepsystscalarlinquadpartlinelindexindexother}
{\Ysssepsystscalarlinquadpartlinelnoindex_{\Ysswayindex \Ysswayindexother}}

\newcommand{\ysssepsystscalarlinquadpartquadelnoindex}
{$\Ysssepsystscalarlinquadpartquadelnoindex$}
\newcommand{\Ysssepsystscalarlinquadpartquadelnoindex}
{q}

\newcommand{\ysssepsystscalarlinquadpartquadelone}
{$\Ysssepsystscalarlinquadpartquadelone$}
\newcommand{\Ysssepsystscalarlinquadpartquadelone}
{\Ysssepsystscalarlinquadpartquadelnoindex_{1}}

\newcommand{\ysssepsystscalarlinquadpartquadeltwo}
{$\Ysssepsystscalarlinquadpartquadeltwo$}
\newcommand{\Ysssepsystscalarlinquadpartquadeltwo}
{\Ysssepsystscalarlinquadpartquadelnoindex_{2}}

\newcommand{\ysssepsystscalarlinquadpartquadelindex}
{$\Ysssepsystscalarlinquadpartquadelindex$}
\newcommand{\Ysssepsystscalarlinquadpartquadelindex}
{\Ysssepsystscalarlinquadpartquadelnoindex_{\Ysswayindex}}



\newcommand{\ysscomplfiltcoefnoindex}
           {$\Ysscomplfiltcoefnoindex$}
\newcommand{\Ysscomplfiltcoefnoindex}
           {h}

\newcommand{\ysscomplfiltcoefindexone}
           {$\Ysscomplfiltcoefindexone$}
\newcommand{\Ysscomplfiltcoefindexone}
           {\Ysscomplfiltcoefnoindex _{1}}

\newcommand{\ysscomplfiltcoefindextwo}
           {$\Ysscomplfiltcoefindextwo$}
\newcommand{\Ysscomplfiltcoefindextwo}
           {\Ysscomplfiltcoefnoindex _{2}}

\newcommand{\ysscomplfiltcoefindexindex}
           {$\Ysscomplfiltcoefindexindex$}
\newcommand{\Ysscomplfiltcoefindexindex}
           {\Ysscomplfiltcoefnoindex _{\Ysswayindex}}

\newcommand{\ysscomplfiltcoefindexindexthird}
           {$\Ysscomplfiltcoefindexindexthird$}
\newcommand{\Ysscomplfiltcoefindexindexthird}
           {\Ysscomplfiltcoefnoindex _{\Ysswayindex \Ysswayindexthird}}

\newcommand{\ysscomplfiltcoefindexotherindexthird}
           {$\Ysscomplfiltcoefindexotherindexthird$}
\newcommand{\Ysscomplfiltcoefindexotherindexthird}
           {\Ysscomplfiltcoefnoindex _{\Ysswayindexother \Ysswayindexthird}}



\newcommand{\ysssepsystadaptgainnoindex}
           {$\Ysssepsystadaptgainnoindex$}
\newcommand{\Ysssepsystadaptgainnoindex}
           {\mu}

\newcommand{\ysssepsystadaptgainone}
           {$\Ysssepsystadaptgainone$}
\newcommand{\Ysssepsystadaptgainone}
           {\Ysssepsystadaptgainnoindex _{1}}

\newcommand{\ysssepsystadaptgaintwo}
           {$\Ysssepsystadaptgaintwo$}
\newcommand{\Ysssepsystadaptgaintwo}
           {\Ysssepsystadaptgainnoindex _{2}}

\newcommand{\ysssepsystadaptgainindex}
           {$\Ysssepsystadaptgainindex$}
\newcommand{\Ysssepsystadaptgainindex}
           {\Ysssepsystadaptgainnoindex _{\Ysswayindex}}



\newcommand{\yodedisctimeval}
          {$\Yodedisctimeval$}
\newcommand{\Yodedisctimeval}
           {\Yssdisctimeval}


\newcommand{\yodesystemstatenoindex}
          {$\Yodesystemstatenoindex$}
\newcommand{\Yodesystemstatenoindex}
           {\theta}

\newcommand{\yodesystemstateindex}
          {$\Yodesystemstateindex$}
\newcommand{\Yodesystemstateindex}
           {\Yodesystemstatenoindex _{\Yodedisctimeval}}

\newcommand{\yodesystemstateeq}
          {$\Yodesystemstateeq$}
\newcommand{\Yodesystemstateeq}
           {\Yodesystemstatenoindex ^{*}}

\newcommand{\yodesystemstatesep}
          {$\Yodesystemstatesep$}
\newcommand{\Yodesystemstatesep}
           {\Yodesystemstatenoindex ^{s}}


\newcommand{\yodesystemsignoindex}
          {$\Yodesystemsignoindex$}
\newcommand{\Yodesystemsignoindex}
           {\xi}

\newcommand{\yodesystemsigtimeindexplusone}
          {$\Yodesystemsigtimeindexplusone$}
\newcommand{\Yodesystemsigtimeindexplusone}
           {\Yodesystemsignoindex _{\Yodedisctimeval + 1}}


\newcommand{\yodesystemfuncnoindex}
          {$\Yodesystemfuncnoindex$}
\newcommand{\Yodesystemfuncnoindex}
           {H}

\newcommand{\yodesystemfuncindex}
          {$\Yodesystemfuncindex$}
\newcommand{\Yodesystemfuncindex}
           {\Yodesystemfuncnoindex ( \Yodesystemstateindex , \Yodesystemsigtimeindexplusone) }


\newcommand{\yodesystemjacobmatrix}
          {$\Yodesystemjacobmatrix$}
\newcommand{\Yodesystemjacobmatrix}
           {J}

\newcommand{\yodesystemjacobmatrixstatesep}
          {$\Yodesystemjacobmatrixstatesep$}
\newcommand{\Yodesystemjacobmatrixstatesep}
           {\Yodesystemjacobmatrix ( \Yodesystemstatesep )}

\newcommand{\yodesystemjacobmatrixlineindex}
          {$\Yodesystemjacobmatrixlineindex$}
\newcommand{\Yodesystemjacobmatrixlineindex}
           {i}

\newcommand{\yodesystemjacobmatrixcolumnindex}
          {$\Yodesystemjacobmatrixcolumnindex$}
\newcommand{\Yodesystemjacobmatrixcolumnindex}
           {j}

\newcommand{\yodesystemjacobmatrixellineindexcolumnindex}
          {$\Yodesystemjacobmatrixellineindexcolumnindex$}
\newcommand{\Yodesystemjacobmatrixellineindexcolumnindex}
           {\Yodesystemjacobmatrix _{\Yodesystemjacobmatrixlineindex \Yodesystemjacobmatrixcolumnindex}}

%
%
%
%
%
%
%
%
\newcommand{\ywritereadonestatenb}
{$\Ywritereadonestatenb$}
\newcommand{\Ywritereadonestatenb}
{K}
%
\newcommand{\ywritereadonestatestepindex}
{$\Ywritereadonestatestepindex$}
\newcommand{\Ywritereadonestatestepindex}
{k}
%
%
%
%
\newcommand{\ytwoqubitseqindex}
{$\Ytwoqubitseqindex$}
\newcommand{\Ytwoqubitseqindex}
{n}
%
%
%
%
%
\newcommand{\yonequbittimewrite}
{$\Yonequbittimewrite$}
\newcommand{\Yonequbittimewrite}
{t_w}
%
\newcommand{\yonequbittimeread}
{$\Yonequbittimeread$}
\newcommand{\Yonequbittimeread}
{t_r}
%
%
%
%
\newcommand{\yonequbittimewritestepindex}
{$\Yonequbittimewritestepindex$}
\newcommand{\Yonequbittimewritestepindex}
{\Yonequbittimewrite (\Ywritereadonestatestepindex)}
%
\newcommand{\yonequbittimereadstepindex}
{$\Yonequbittimereadstepindex$}
\newcommand{\Yonequbittimereadstepindex}
{\Yonequbittimeread (\Ywritereadonestatestepindex)}
%
%
\newcommand{\ytwoqubitwritereadtimeinterval}
{$\Ytwoqubitwritereadtimeinterval$}
\newcommand{\Ytwoqubitwritereadtimeinterval}
{T}
%
\newcommand{\ytwoqubitwritereadtimeintervalindex}
{$\Ytwoqubitwritereadtimeintervalindex$}
\newcommand{\Ytwoqubitwritereadtimeintervalindex}
{\Ytwoqubitwritereadtimeinterval (\Ywritereadonestatestepindex)}
%
%
%
%
\newcommand{\ytwoqubitsbasisplusplus}
{$\Ytwoqubitsbasisplusplus$}
\newcommand{\Ytwoqubitsbasisplusplus}
{{\cal B} _+}
%
\newcommand{\ytwoqubitsbasisoneone}
{$\Ytwoqubitsbasisoneone$}
\newcommand{\Ytwoqubitsbasisoneone}
{{\cal B} _1}
%
%
%
%
\newcommand{\ytwoqubitsprobaplusplus}
{$\Ytwoqubitsprobaplusplus$}
\newcommand{\Ytwoqubitsprobaplusplus}
{p_1}
%
\newcommand{\ytwoqubitsprobaminusminus}
{$\Ytwoqubitsprobaminusminus$}
\newcommand{\Ytwoqubitsprobaminusminus}
{p_2}
%
\newcommand{\ytwoqubitsprobaplusminus}
{$\Ytwoqubitsprobaplusminus$}
\newcommand{\Ytwoqubitsprobaplusminus}
{p_3}
%
\newcommand{\ytwoqubitsprobaminusplus}
{$\Ytwoqubitsprobaminusplus$}
\newcommand{\Ytwoqubitsprobaminusplus}
{p_4}
%
\newcommand{\ytwoqubitsprobaindexstd}
{$\Ytwoqubitsprobaindexstd$}
\newcommand{\Ytwoqubitsprobaindexstd}
{p_j}

%
%
%
%
\newcommand{\yqubitindexstd}
{$\Yqubitindexstd$}
\newcommand{\Yqubitindexstd}
{i}
%
%
%
%
\newcommand{\yparamqubitbothstateplusmodulusnot}
{$\Yparamqubitbothstateplusmodulusnot$}
\newcommand{\Yparamqubitbothstateplusmodulusnot}
{r}
%
\newcommand{\yparamqubitonestateplusmodulus}
{$\Yparamqubitonestateplusmodulus$}
\newcommand{\Yparamqubitonestateplusmodulus}
{{\Yparamqubitbothstateplusmodulusnot}_1}
%
\newcommand{\yparamqubittwostateplusmodulus}
{$\Yparamqubittwostateplusmodulus$}
\newcommand{\Yparamqubittwostateplusmodulus}
{{\Yparamqubitbothstateplusmodulusnot}_2}
%
\newcommand{\yparamqubitindexstdstateplusmodulus}
{$\Yparamqubitindexstdstateplusmodulus$}
\newcommand{\Yparamqubitindexstdstateplusmodulus}
{{\Yparamqubitbothstateplusmodulusnot}_{\Yqubitindexstd}}
%
\newcommand{\yparamqubitbothstateminusmodulusnot}
{$\Yparamqubitbothstateminusmodulusnot$}
\newcommand{\Yparamqubitbothstateminusmodulusnot}
{q}
%
\newcommand{\yparamqubitonestateminusmodulus}
{$\Yparamqubitonestateminusmodulus$}
\newcommand{\Yparamqubitonestateminusmodulus}
{{\Yparamqubitbothstateminusmodulusnot}_1}
%
\newcommand{\yparamqubittwostateminusmodulus}
{$\Yparamqubittwostateminusmodulus$}
\newcommand{\Yparamqubittwostateminusmodulus}
{{\Yparamqubitbothstateminusmodulusnot}_2}
%
\newcommand{\yparamqubitindexstdstateminusmodulus}
{$\Yparamqubitindexstdstateminusmodulus$}
\newcommand{\Yparamqubitindexstdstateminusmodulus}
{{\Yparamqubitbothstateminusmodulusnot}_{\Yqubitindexstd}}
%
\newcommand{\yparamqubitonestateplusphase}
{$\Yparamqubitonestateplusphase$}
\newcommand{\Yparamqubitonestateplusphase}
{\theta_1}
%
\newcommand{\yparamqubittwostateplusphase}
{$\Yparamqubittwostateplusphase$}
\newcommand{\Yparamqubittwostateplusphase}
{\theta_2}
%
\newcommand{\yparamqubitonestateminusphase}
{$\Yparamqubitonestateminusphase$}
\newcommand{\Yparamqubitonestateminusphase}
{\phi_1}
%
\newcommand{\yparamqubittwostateminusphase}
{$\Yparamqubittwostateminusphase$}
\newcommand{\Yparamqubittwostateminusphase}
{\phi_2}
%
%
%
%
\newcommand{\ytwoqubitresultphaseinit}
{$\Ytwoqubitresultphaseinit$}
\newcommand{\Ytwoqubitresultphaseinit}
{\Delta _I}
%
\newcommand{\ytwoqubitresultphaseevol}
{$\Ytwoqubitresultphaseevol$}
\newcommand{\Ytwoqubitresultphaseevol}
{\Delta _E}
%
\newcommand{\ytwoqubitresultphaseevolsin}
{$\Ytwoqubitresultphaseevolsin$}
\newcommand{\Ytwoqubitresultphaseevolsin}
{v}
%
%
%
%
\newcommand{\ymagfieldnot}
{$\Ymagfieldnot$}
\newcommand{\Ymagfieldnot}
{B}
%
%
%
\newcommand{\yhamiltonfieldscale}
{$\Yhamiltonfieldscale$}
\newcommand{\Yhamiltonfieldscale}
{G}
%
%
%
%
\newcommand{\ymixfunc}
{$\Ymixfunc$}
\newcommand{\Ymixfunc}
{g}
%
%
%
%
\newcommand{\ymixfuncexponent}
{$\Ymixfuncexponent$}
\newcommand{\Ymixfuncexponent}
{k}
%
%
%
%
\newcommand{\ymixfuncimpldiff}
{$\Ymixfuncimpldiff$}
\newcommand{\Ymixfuncimpldiff}
{F}
%
%
%
\newcommand{\ymixfuncjacob}
{$\Ymixfuncjacob$}
\newcommand{\Ymixfuncjacob}
{J_{\Ymixfunc}}
%
\newcommand{\ymixfuncjacobval}
{$\Ymixfuncjacobval$}
\newcommand{\Ymixfuncjacobval}
{\Ymixfuncjacob ( \Ysssrcsigdisctimecentvec )}
%
%
%
%
\newcommand{\yparamqubitindexstdstateplusmodulussign}
{$\Yparamqubitindexstdstateplusmodulussign$}
\newcommand{\Yparamqubitindexstdstateplusmodulussign}
{\epsilon}
%
%
%
%
%
\newcommand{\ysssrcsigdisctimecentvecrand}
{$\Ysssrcsigdisctimecentvecrand$}
\newcommand{\Ysssrcsigdisctimecentvecrand}
{S}
%
\newcommand{\ysssrcsigdisctimecentindexrand}
{$\Ysssrcsigdisctimecentindexrand$}
\newcommand{\Ysssrcsigdisctimecentindexrand}
{\Ysssrcsigdisctimecentvecrand _ {\Ysswayindex}}
%
%
%
\newcommand{\ysssrcsigdisctimecentvecranddens}
{$\Ysssrcsigdisctimecentvecranddens$}
\newcommand{\Ysssrcsigdisctimecentvecranddens}
{f_{\Ysssrcsigdisctimecentvecrand}}
%
\newcommand{\ysssrcsigdisctimecentvecranddensval}
{$\Ysssrcsigdisctimecentvecranddensval$}
\newcommand{\Ysssrcsigdisctimecentvecranddensval}
{\Ysssrcsigdisctimecentvecranddens ( \Ysssrcsigdisctimecentvec )}
%
\newcommand{\ysssrcsigdisctimecentindexranddens}
{$\Ysssrcsigdisctimecentindexranddens$}
\newcommand{\Ysssrcsigdisctimecentindexranddens}
{f_{\Ysssrcsigdisctimecentindexrand}}
%
\newcommand{\ysssrcsigdisctimecentindexranddensval}
{$\Ysssrcsigdisctimecentindexranddensval$}
\newcommand{\Ysssrcsigdisctimecentindexranddensval}
{\Ysssrcsigdisctimecentindexranddens ( \Ysssrcsigdisctimecentindex )}
%
%
%
\newcommand{\yssmixsigdisctimecentvecrand}
{$\Yssmixsigdisctimecentvecrand$}
\newcommand{\Yssmixsigdisctimecentvecrand}
{X}
%
\newcommand{\yssmixsigdisctimecentvecranddens}
{$\Yssmixsigdisctimecentvecranddens$}
\newcommand{\Yssmixsigdisctimecentvecranddens}
{f_{\Yssmixsigdisctimecentvecrand}}
%
\newcommand{\yssmixsigdisctimecentvecranddensval}
{$\Yssmixsigdisctimecentvecranddensval$}
\newcommand{\Yssmixsigdisctimecentvecranddensval}
{\Yssmixsigdisctimecentvecranddens ( \Yssmixsigdisctimecentvec )}
%
%
%
%
\newcommand{\ymlsampnb}
{$\Ymlsampnb$}
\newcommand{\Ymlsampnb}
{M}
%
%
%
\newcommand{\ymlcost}
{$\Ymlcost$}
\newcommand{\Ymlcost}
{L}
%
%
%
\newcommand{\ymlcostlnmean}
{$\Ymlcostlnmean$}
\newcommand{\Ymlcostlnmean}
{{\cal L}}
%
\newcommand{\ymlcostlnmeanargmixparam}
{$\Ymlcostlnmeanargmixparam$}
\newcommand{\Ymlcostlnmeanargmixparam}
{\Ymlcostlnmean ( \Ytwoqubitresultphaseevolsin ,
                  \Ysssrcsigdisctimecentone (\Ytwoqubitresultphaseevolsin),
                  \Ysssrcsigdisctimecenttwo (\Ytwoqubitresultphaseevolsin),
                  \Ysssrcsigdisctimecentthree (\Ytwoqubitresultphaseevolsin)
                )}

%
%
%
%
%
%
\newcommand{\ysqrtminusone}
{$\Ysqrtminusone$}
\newcommand{\Ysqrtminusone}
{i}
%
%
%

%
%
%
%
%
%
\newcommand{\ysepfunc}
{$\Ysepfunc$}
\newcommand{\Ysepfunc}
{h}
%
%
%
\newcommand{\ysepfuncjacob}
{$\Ysepfuncjacob$}
\newcommand{\Ysepfuncjacob}
{J_{\Ysepfunc}}
%
%
%
\newcommand{\ysssepsystmatrixscalareloptim}
          {$\Ysssepsystmatrixscalareloptim$}
\newcommand{\Ysssepsystmatrixscalareloptim}
           {\Ysssepsystmatrixscalarelnoindex _{k \ell}}


\newcommand{\yetwelveinfoRVcontentropydiff}
{$\YetwelveinfoRVcontentropydiff$}
\newcommand{\YetwelveinfoRVcontentropydiff}
{H}
%


\newcommand{\yetwelveinfoinfo}{$\Yetwelveinfoinfo$}
\newcommand{\Yetwelveinfoinfo}{I}

\newcommand{\yetwelveinfoinfominusconst}{$\Yetwelveinfoinfominusconst$}
\newcommand{\Yetwelveinfoinfominusconst}{C}
%
\newcommand{\yetwelveinfoinfominusconstargestimparam}
{$\Yetwelveinfoinfominusconstargestimparam$}
\newcommand{\Yetwelveinfoinfominusconstargestimparam}
{\Yetwelveinfoinfominusconst
(
\Ysssepsystmatrixscalarelonetwo ,
\Ysssepsystmatrixscalareltwoone ,
\ytextblue{\Yssoutsepsystsigdisctimecentonerand}
(
\Ysssepsystmatrixscalarelonetwo ,
\Ysssepsystmatrixscalareltwoone
),
\ytextblue{\Yssoutsepsystsigdisctimecenttworand}
(
\Ysssepsystmatrixscalarelonetwo ,
\Ysssepsystmatrixscalareltwoone
)
)
}


%
%
%


%
\newcommand{\yssoutsepsystsigdisctimecentvecrand}
{$\Yssoutsepsystsigdisctimecentvecrand$}
\newcommand{\Yssoutsepsystsigdisctimecentvecrand}
{Y}
%
\newcommand{\yssoutsepsystsigdisctimecentonerand}
{$\Yssoutsepsystsigdisctimecentonerand$}
\newcommand{\Yssoutsepsystsigdisctimecentonerand}
{\Yssoutsepsystsigdisctimecentvecrand _ {1}}
%
\newcommand{\yssoutsepsystsigdisctimecenttworand}
{$\Yssoutsepsystsigdisctimecenttworand$}
\newcommand{\Yssoutsepsystsigdisctimecenttworand}
{\Yssoutsepsystsigdisctimecentvecrand _ {2}}
%
\newcommand{\yssoutsepsystsigdisctimecentindexrand}
{$\Yssoutsepsystsigdisctimecentindexrand$}
\newcommand{\Yssoutsepsystsigdisctimecentindexrand}
{\Yssoutsepsystsigdisctimecentvecrand _ {\Ysswayindex}}
%
%
%
\newcommand{\yssoutsepsystsigdisctimecentindexranddens}
{$\Yssoutsepsystsigdisctimecentindexranddens$}
\newcommand{\Yssoutsepsystsigdisctimecentindexranddens}
{f_{
\ytextblue
{\Yssoutsepsystsigdisctimecentindexrand}
}}
\newcommand{\ytextred}[1]
{{#1}}
\newcommand{\ytextblue}[1]
{{#1}}
\normalsize
\sloppy
\begin{center}
{
\Large
\bf
Effect of indirect dependencies on
"A mutual information minimization approach for a class of nonlinear
recurrent separating systems"
}
~\\
~\\
Yannick Deville$^1$,
Alain Deville$^2$,
and
Shahram Hosseini$^1$
~\\
~\\
(1) 
Laboratoire
d'Astrophysique de Toulouse-Tarbes,
Universit\'e de Toulouse, CNRS,
14 Av. Edouard Belin, 31400 Toulouse, France.
Email:
ydeville@ast.obs-mip.fr ,
shosseini@ast.obs-mip.fr
~\\
(2) 
IM2NP, Universit\'e de Provence, Centre de
Saint-J\'er\^ome, 13397 Marseille Cedex 20, France.
Email: alain.deville@univ-provence.fr
\end{center}
~\\
~\\
~\\
{\bf Abstract.}
In a recent paper
\cite{article-lduarte-mlsp2007},
Duarte and Jutten
investigated the Blind Source Separation (BSS) problem, for
the nonlinear mixing model that they introduced 
in that paper.
They proposed to solve this problem by using 
information-theoretic 
tools, more precisely by minimizing
the mutual information (MI) of the outputs of
the separating structure.
When 
applying
the MI approach to BSS problems,
one usually 
determines
the analytical expressions of
the 
derivatives
of the MI with respect to the parameters of
the considered separating model.
In the literature,
these calculations
were mainly 
reported
for 
linear mixtures
up to now. They are 
more complex for nonlinear mixtures,
due to 
dependencies 
between the considered quantities.
Moreover, the notations commonly employed by the BSS community in such
calculations may become misleading when using them for
nonlinear mixtures, due to the above-mentioned dependencies.
We claim that the calculations reported in
\cite{article-lduarte-mlsp2007}
contain an error, because they did not take into account
all these dependencies.
In this document, we therefore explain this phenomenon, by showing
the effect of indirect dependencies
on the application of the MI approach to the mixing 
\ytextblue{and separating models}
considered in
\cite{article-lduarte-mlsp2007}.
We thus introduce
a corrected expression of the
gradient of the considered 
BSS criterion
based on MI.
This correct gradient
may then e.g.
be used to optimize
the adaptive coefficients
of the considered
separating system by means of the well-known gradient descent
algorithm.
As explained 
hereafter,
this investigation
has some similarities with an analysis that we previously
reported in another arXiv 
document
\cite{bibref-ellipses-chap-mv-moi-arxiv}.
However,
these two investigations concern different problems,
not only in terms of the considered type of 
\ytextblue{mixture and separating structure},
but also of the mathematical tools used to
develop BSS methods for these 
\ytextblue{configurations}
(information theory vs maximum likelihood approach).
~\\
~\\
{\bf Keywords.}
Information theory,
mutual information,
blind signal separation, 
independent component analysis,
nonlinear \ytextblue{mixture,
additive-target mixture (ATM),
recurrent separating structure},
indirect dependency,
total derivative,
partial derivative,
gradient.
~\\
~\\
\section{%
Data model}
Blind
source separation (BSS) consists in restoring a vector 
\ysssrcsigconttimecentvecval\
of
\ysssrcnb\
unknown
source signals from a vector
\yssmixsigconttimecentvecval\
of
\ysssensnb\
observed signals
(most often with
$
\Ysssensnb
=
\Ysssrcnb
$),
where
\yssmixsigconttimecentvecval\
is derived from
\ysssrcsigconttimecentvecval\
through an unknown mixing function \ymixfunc , 
i.e.
\begin{equation}
\label{eq-mix-model-with-time}
\Yssmixsigconttimecentvecval
=
\Ymixfunc
(
\Ysssrcsigconttimecentvecval
) .
\end{equation}
Recently, 
Duarte and 
Jutten
investigated a specific version of this problem
\cite{article-lduarte-mlsp2007},
which involves
$
\Ysssensnb
=
2
$
observed signals
$
\Yssmixsigconttimecentoneval
$
and
$
\Yssmixsigconttimecenttwoval
$,
which are derived
from
$
\Ysssrcnb
=
2
$
source signals 
$
\Ysssrcsigconttimecentoneval
$
and
$
\Ysssrcsigconttimecenttwoval
$,
through the nonlinear function
defined as
\begin{eqnarray}
\label{eq-mixsigconttimecentoneval}
\Yssmixsigconttimecentoneval
&
=
&
\Ysssrcsigconttimecentoneval
+
\Yssmixmatrixscalarelonetwo
(\Ysssrcsigconttimecenttwoval)
^{\Ymixfuncexponent}
\\
\label{eq-mixsigconttimecenttwoval}
\Yssmixsigconttimecenttwoval
&
=
&
\Ysssrcsigconttimecenttwoval
+
\Yssmixmatrixscalareltwoone
(\Ysssrcsigconttimecentoneval)
^{
\frac{1}{\Ymixfuncexponent}
}
.
\end{eqnarray}
This data model is derived from
the Nikolsky-Eisenman empirical
model
for potentiometric-based ion concentration
sensors
\cite{article-lduarte-mlsp2007}.
As in
\cite{article-lduarte-mlsp2007},
we omit the time index
$
t
$
in signal notations hereafter, for readability.
The mixing model 
(\ref{eq-mixsigconttimecentoneval})-%
(\ref{eq-mixsigconttimecenttwoval})
may then also be expressed in
compact form as
\begin{equation}
\label{eq-mix-model-vector}
\Yssmixsigdisctimecentvec
=
\Ymixfunc
(
\Ysssrcsigdisctimecentvec
)
.
\end{equation}
In this equation,
$
\Ysssrcsigdisctimecentvec
=
[
\Ysssrcsigconttimecentone ,
\Ysssrcsigconttimecenttwo
]
^{T}
$
and
$
\Yssmixsigdisctimecentvec
=
[
\Yssmixsigconttimecentone ,
\Yssmixsigconttimecenttwo
]
^{T}
$,
where 
$
^{T}
$
stands for transpose,
and
the
nonlinear mixing function
\ymixfunc\
has
two
components
$
\Ymixfunc _{1}
$
and
$
\Ymixfunc _{2}
$,
with
$
\Yssmixsigdisctimecentindex
=
\Ymixfunc _{\Ysswayindex}
(
\Ysssrcsigdisctimecentvec
),
$
$
\
\forall 
\Ysswayindex \in \{ 1, 2 \}
$.
These 
components 
$
\Ymixfunc _{\Ysswayindex}
$
are
respectively defined by
(\ref{eq-mixsigconttimecentoneval})
and
(\ref{eq-mixsigconttimecenttwoval}).
Eq.
(\ref{eq-mix-model-vector}) focuses on the signals (i.e. sources
and observations).
It hides the fact that the observations also depend on the
parameters of the mixing model, i.e. on 
\yssmixmatrixscalarelonetwo\
and
\yssmixmatrixscalareltwoone\
in the model considered here.
This additional dependency can be made explicit, by rewriting
(\ref{eq-mix-model-vector}) as
\begin{equation}
\label{eq-mix-model-vector-with-mix-param}
\Yssmixsigdisctimecentvec
=
\Ymixfunc
(
\Ysssrcsigdisctimecentvec ,
\Yssmixmatrixscalarelonetwo ,
\Yssmixmatrixscalareltwoone)
.
\end{equation}
\section{Previously reported results for mutual information
minimization}
\label{sec-ml-previous}
\subsection{Overview and issue of previous method}
As suggested above,
the BSS problem associated 
with
the mixing model
(\ref{eq-mixsigconttimecentoneval})-%
(\ref{eq-mixsigconttimecenttwoval})
consists in
retrieving a sequence of
unknown source vectors 
$
\Ysssrcsigdisctimecentvec
$
from the corresponding sequence
of measured observation vectors
$
\Yssmixsigdisctimecentvec
$
and from
the 
mixing parameters
\yssmixmatrixscalarelonetwo\
and
\yssmixmatrixscalareltwoone ,
which
are also initially unknown.
These mixing parameters
should therefore be estimated before proceeding to the source
restoration step.
Creating an overall BSS method thus consists in defining
two items, i.e. i) a "separating structure", which performs the
inversion of the mixing equations
(\ref{eq-mixsigconttimecentoneval})-%
(\ref{eq-mixsigconttimecenttwoval})
for known mixing parameter values, and ii) a procedure
for estimating these mixing parameters.

The separating structure used
in
\cite{article-lduarte-mlsp2007}
was derived by
Duarte and Jutten
from the 
structure for linear-quadratic mixtures
proposed by Hosseini and
Deville in
\cite{amoi4-39},%
\cite{amoi4-48},%
\cite{amoi5-32},%
\cite{amoi5-49}.
The structure in
\cite{article-lduarte-mlsp2007}
belongs to the general
class of structures proposed by
Deville and Hosseini in 
\cite{amoi5-49}
for
the
ATM class of mixing models,
which includes the specific model
(\ref{eq-mixsigconttimecentoneval})-%
(\ref{eq-mixsigconttimecenttwoval}).

As for
the estimation of the mixing parameters,
Duarte and Jutten
developed a procedure based on 
information-theoretic 
tools, more precisely on the minimization
of the mutual information (MI) of the outputs of
the separating structure.
However,
we here claim that this 
procedure
contains an
error, 
which is due to a difficulty encountered
with \emph{nonlinear} 
mixing models in general, for different classes of BSS methods.
This difficulty
is somewhat similar to
the one
that we highlighted 
in another arXiv 
document
\cite{bibref-ellipses-chap-mv-moi-arxiv}:
unlike the method considered hereafter,
the BSS approach described in
\cite{bibref-ellipses-chap-mv-moi-arxiv}
is not based on information theoretic tools,
but on the maximum likelihood framework.
Moreover, it concerns a different class of nonlinear mixtures.
However, 
similar quantities
appear in the
calculations performed for
both methods\footnote{The quantities to be 
respectively considered in these two methods depend on
different
signals (source signals vs outputs of separating system)
and functions 
(mixing function vs separating function).
However, these signals and functions yield similar phenomena
concerning the topic addressed in this document.},
and they deserve special care in both of
them.

The current document therefore aims at explaining
and correcting the error which was made in 
\cite{article-lduarte-mlsp2007}.
We thus show how the BSS method of
\cite{article-lduarte-mlsp2007}
should be modified so as to
actually achieve mutual information
minimization.
Before 
focusing on the issue faced in
\cite{article-lduarte-mlsp2007},
we now summarize the features of 
that approach
which are of importance
hereafter.

\subsection{Description 
of previous method}
The considered separating structure
has internal
adaptive coefficients
\ysssepsystmatrixscalarelonetwo\
and
\ysssepsystmatrixscalareltwoone .
For each time 
$
t
$,
this structure
determines and output vector
$
\Yssoutsepsystsigconttimecentvec
=
[
\Yssoutsepsystsigdisctimecentone
,
\Yssoutsepsystsigdisctimecenttwo
 ] ^{T}
$
from its current internal coefficients
and from
the current observation vector
$
\Yssmixsigdisctimecentvec
$.
To this end, it iteratively updates its output
according to
\begin{eqnarray}
\Yssoutsepsystsigdisctimecentone
(n + 1)
&
=
&
\Yssmixsigdisctimecentone
-
\Ysssepsystmatrixscalarelonetwo
(
\Yssoutsepsystsigdisctimecenttwo
(n )
)
^{
\Ymixfuncexponent
}
\\
\Yssoutsepsystsigdisctimecenttwo
(n + 1)
&
=
&
\Yssmixsigdisctimecenttwo
-
\Ysssepsystmatrixscalareltwoone
(
\Yssoutsepsystsigdisctimecentone
(n )
)
^{
\frac{1}{\Ymixfuncexponent}
}
.
\end{eqnarray}
The convergence of this recurrence therefore 
corresponds to a state such that
\begin{eqnarray}
\label{eq-outsepsystsigdisctimecentone-converg}
\Yssoutsepsystsigdisctimecentone
&
=
&
\Yssmixsigdisctimecentone
-
\Ysssepsystmatrixscalarelonetwo
\Yssoutsepsystsigdisctimecenttwo
^{
\Ymixfuncexponent
}
\\
\label{eq-outsepsystsigdisctimecenttwo-converg}
\Yssoutsepsystsigdisctimecenttwo
&
=
&
\Yssmixsigdisctimecenttwo
-
\Ysssepsystmatrixscalareltwoone
\Yssoutsepsystsigdisctimecentone
^{
\frac{1}{\Ymixfuncexponent}
}
.
\end{eqnarray}
\ytextblue{For a given time
$
t
$,
we 
denote as
\yssoutsepsystsigdisctimecentonerand\
and
\yssoutsepsystsigdisctimecenttworand\
the random variables respectively
associated with the output signal
samples
\yssoutsepsystsigdisctimecentone\
and
\yssoutsepsystsigdisctimecenttwo\ obtained
after the above recurrence has converged.
We also define the corresponding output random
vector as
$
\Yssoutsepsystsigdisctimecentvecrand
=
[
\Yssoutsepsystsigdisctimecentonerand
,
\Yssoutsepsystsigdisctimecenttworand
 ] ^{T}
$.}

The optimum values of
\ysssepsystmatrixscalarelonetwo\
and
\ysssepsystmatrixscalareltwoone\
are defined as those which minimize
the mutual information 
of
\ytextblue{\yssoutsepsystsigdisctimecentonerand\
and
\yssoutsepsystsigdisctimecenttworand ,}
which is denoted
$
\Yetwelveinfoinfo
(
\ytextblue{\Yssoutsepsystsigdisctimecentvecrand}
)
$.
Equivalently, they are those which minimize a
quantity
$
\Yetwelveinfoinfominusconst
(
\ytextblue{\Yssoutsepsystsigdisctimecentvecrand}
)
$.
This quantity is equal to
$
\Yetwelveinfoinfo
(
\ytextblue{\Yssoutsepsystsigdisctimecentvecrand}
)
$,
up to an additive term which 
only depends on the observations and which
therefore
does not depend
on
\ysssepsystmatrixscalarelonetwo\
and
\ysssepsystmatrixscalareltwoone .
That quantity reads
\begin{equation}
\label{eq-ln-pdf-mix-vs-src-versiontwo}
\Yetwelveinfoinfominusconst
(
\ytextblue{\Yssoutsepsystsigdisctimecentvecrand}
)
=
\left(
\sum_{\Ysswayindex = 1}^{2}
\YetwelveinfoRVcontentropydiff
(
\ytextblue{\Yssoutsepsystsigdisctimecentindexrand}
)
\right)
-
E \{
\ln
|
\Ysepfuncjacob
|
\}
\end{equation}
where
$
\YetwelveinfoRVcontentropydiff
(
\ytextblue{\Yssoutsepsystsigdisctimecentindexrand}
)
$
is the differential entropy of
\ytextblue{\yssoutsepsystsigdisctimecentindexrand}
while
$E \{ . \}$
stands for expectation
and
\ysepfuncjacob\
is the
Jacobian\footnote{\ytextblue{For the sake of readability, we use the same notation,
i.e. \ysepfuncjacob , for (i) the sample value of this Jacobian
associated to sample values
\yssoutsepsystsigdisctimecentone\
and
\yssoutsepsystsigdisctimecenttwo\
(see e.g.
(\ref{eq-mixfuncjacobval})) and
(ii) the random variable defined by this quantity when considered as
a function of the random variables
\yssoutsepsystsigdisctimecentonerand\
and
\yssoutsepsystsigdisctimecenttworand\
(see e.g. 
(\ref{eq-costlnmean-gradient})).
To know whether we are considering the sample value of
\ysepfuncjacob\ or the associated random variable
in an equation, one just
has to 
check whether that equation involves the
sample values
\yssoutsepsystsigdisctimecentone\
and
\yssoutsepsystsigdisctimecenttwo\
or the associated
random variables
\yssoutsepsystsigdisctimecentonerand\
and
\yssoutsepsystsigdisctimecenttworand :
see 
e.g.
(\ref{eq-mixfuncjacobval}) and
(\ref{eq-costlnmean-gradient}).
}
}
of the 
separating function
$
\Ysepfunc \
\ytextred{
=
\Ymixfunc ^{-1}
}
$
achieved by the considered separating structure,
i.e. \ysepfuncjacob\
is the determinant of the Jacobian matrix of \ysepfunc .
For the function
\ysepfunc\ considered in this investigation, 
the authors show that
\begin{equation}
\label{eq-mixfuncjacobval}
\Ysepfuncjacob
=
\frac{1}
{
1
-
\Ysssepsystmatrixscalarelonetwo
\Ysssepsystmatrixscalareltwoone
\Yssoutsepsystsigdisctimecentone
^{
\frac{1}{\Ymixfuncexponent} - 1
}
\Yssoutsepsystsigdisctimecenttwo
^{
\Ymixfuncexponent -1
}
}
.
\end{equation}

To determine the
values of
\ysssepsystmatrixscalarelonetwo\
and
\ysssepsystmatrixscalareltwoone\
which minimize
$
\Yetwelveinfoinfominusconst
(
\ytextblue{\Yssoutsepsystsigdisctimecentvecrand}
)
$,
the authors then consider the gradient of
$
\Yetwelveinfoinfominusconst
(
\ytextblue{\Yssoutsepsystsigdisctimecentvecrand}
)
$
with respect to the vector 
composed of
\ysssepsystmatrixscalarelonetwo\
and
\ysssepsystmatrixscalareltwoone .
Each
component of this gradient is equal to
the derivative
of
$
\Yetwelveinfoinfominusconst
(
\ytextblue{\Yssoutsepsystsigdisctimecentvecrand}
)
$
with respect to 
one of the parameters
\ysssepsystmatrixscalareloptim .
In 
\cite{article-lduarte-mlsp2007}, 
the authors denoted this gradient
by using the notation
most often 
employed
in the BSS community (see e.g.
\cite{icabook-oja}),
i.e. 
each of its components reads
$
\frac{\partial 
\Yetwelveinfoinfominusconst
(
\ytextblue{\Yssoutsepsystsigdisctimecentvecrand}
)
}{\partial \Ysssepsystmatrixscalareloptim}
$.
We keep this notation in this section, in order to clearly refer to
the equations available in
\cite{article-lduarte-mlsp2007},
but in Section
\ref{sec-ml-new} we will show 
that it may be misleading and we will therefore introduce another
notation.
So, in
\cite{article-lduarte-mlsp2007},
it was showed 
that these derivatives read
\begin{equation}
\label{eq-costlnmean-gradient}
\frac{\partial 
\Yetwelveinfoinfominusconst
(
\ytextblue{\Yssoutsepsystsigdisctimecentvecrand}
)
}{\partial \Ysssepsystmatrixscalareloptim}
=
\left(
\sum_{\Ysswayindex = 1}^{2}
E \{
\psi_{\Ysswayindex} (
\ytextblue{\Yssoutsepsystsigdisctimecentindexrand}
)
\frac{\partial
\ytextblue{\Yssoutsepsystsigdisctimecentindexrand}
}{\partial \Ysssepsystmatrixscalareloptim} \}
\right)
-
E \{
\frac{1}{\Ysepfuncjacob}
\frac{
\partial \Ysepfuncjacob
}{
\partial \Ysssepsystmatrixscalareloptim
}
\}
\end{equation}
where
\begin{equation}
\label{eq-score-func-def}
\psi_{\Ysswayindex}(u)
=
-\frac{d \ln{
\Yssoutsepsystsigdisctimecentindexranddens
(u)}}{d u}
\hspace{5mm}
\forall
\Ysswayindex
\in \{ 1 , 2 \}
\end{equation}
are the score functions of the output signals,
denoting
$
\Yssoutsepsystsigdisctimecentindexranddens
( .
)
$
the probability density functions of these signals.

The last stage of this investigation consists
in deriving the expressions of all the terms of
the right-hand side
of
(\ref{eq-costlnmean-gradient}).
In Equation (26) of
\cite{article-lduarte-mlsp2007},
an explicit expression 
is provided 
and it is stated that it is equal
to (the vector form of)
the term
$
E \{
\frac{1}{\Ysepfuncjacob}
\frac{
\partial \Ysepfuncjacob
}{
\partial \Ysssepsystmatrixscalareloptim
}
\}
$
which appears in
(\ref{eq-costlnmean-gradient}).
We claim that this is not true,
because
the expression whose expectation is
provided in the right-hand side of
Equation (26) of
\cite{article-lduarte-mlsp2007}
is \emph{only one of the terms} which compose the
complete expression to be then used in
(\ref{eq-costlnmean-gradient})
as the term 
misleadingly denoted
$
\frac{1}{\Ysepfuncjacob}
\frac{
\partial \Ysepfuncjacob
}{
\partial \Ysssepsystmatrixscalareloptim
}
$
in
(\ref{eq-costlnmean-gradient}).
In the following section of the current document, 
we clarify this point
and we determine
the 
complete expression of the term denoted
$
\frac{1}{\Ysepfuncjacob}
\frac{
\partial \Ysepfuncjacob
}{
\partial \Ysssepsystmatrixscalareloptim
}
$
in
(\ref{eq-costlnmean-gradient}).
We also comment about the other terms of
(\ref{eq-costlnmean-gradient}).
\section{New results for mutual information minimization:
corrected expression of gradient}
\label{sec-ml-new}
When
determining the
values of
\ysssepsystmatrixscalarelonetwo\
and
\ysssepsystmatrixscalareltwoone\
which minimize
$
\Yetwelveinfoinfominusconst
(
\ytextblue{\Yssoutsepsystsigdisctimecentvecrand}
)
$,
that function
$
\Yetwelveinfoinfominusconst
(
\ytextblue{\Yssoutsepsystsigdisctimecentvecrand}
)
$
is considered for the fixed set of observed vectors.
The only independent
variable in this approach is the set of parameters
to be estimated, i.e.
\ysssepsystmatrixscalarelonetwo\
and
\ysssepsystmatrixscalareltwoone .
The outputs 
$
\Yssoutsepsystsigdisctimecentone$
and
$
\Yssoutsepsystsigdisctimecenttwo
$
of the separating system are
dependent variables, 
here linked to the observations and to
\ysssepsystmatrixscalarelonetwo\
and
\ysssepsystmatrixscalareltwoone\
by
(\ref{eq-outsepsystsigdisctimecentone-converg})-%
(\ref{eq-outsepsystsigdisctimecenttwo-converg}).
The overall variations of 
$
\Yetwelveinfoinfominusconst
(
\ytextblue{\Yssoutsepsystsigdisctimecentvecrand}
)
$
with respect to 
\ysssepsystmatrixscalarelonetwo\
and
\ysssepsystmatrixscalareltwoone\
result from 
two types of terms contained in the expression of
$
\Yetwelveinfoinfominusconst
(
\ytextblue{\Yssoutsepsystsigdisctimecentvecrand}
)
$, i.e.
(i) the terms involving 
\ysssepsystmatrixscalarelonetwo\
and
\ysssepsystmatrixscalareltwoone\
themselves and
(ii) the terms involving 
the output 
\ytextblue{random variables
$
\Yssoutsepsystsigdisctimecentonerand$
and
$
\Yssoutsepsystsigdisctimecenttworand
$,}
which are here considered
as functions of
\ysssepsystmatrixscalarelonetwo\
and
\ysssepsystmatrixscalareltwoone\
and which may therefore be denoted as
$
\ytextblue{\Yssoutsepsystsigdisctimecentonerand}
(
\Ysssepsystmatrixscalarelonetwo ,
\Ysssepsystmatrixscalareltwoone
)
$
and
$
\ytextblue{\Yssoutsepsystsigdisctimecenttworand}
(
\Ysssepsystmatrixscalarelonetwo ,
\Ysssepsystmatrixscalareltwoone
)
$
for the sake of clarity.

This approach should be kept in mind when interpreting all equations in
\cite{article-lduarte-mlsp2007},
which were partly gathered in
Section
\ref{sec-ml-previous} of the current document.
Especially, the 
function
$
\Yetwelveinfoinfominusconst
(
\ytextblue{\Yssoutsepsystsigdisctimecentvecrand}
)
$
itself, which appears in the left-hand
side of
(\ref{eq-ln-pdf-mix-vs-src-versiontwo}),
may be denoted as
\yetwelveinfoinfominusconstargestimparam\
for the sake of clarity.
In order to determine the location
of the minimum of
this function,
one should then 
consider
the \emph{total} derivatives of
\yetwelveinfoinfominusconstargestimparam\
with respect to
\ysssepsystmatrixscalarelonetwo\
and
\ysssepsystmatrixscalareltwoone .
The notations with partial derivatives
in
(\ref{eq-costlnmean-gradient})
may therefore be misleading, 
as confirmed below.
Therefore,
(\ref{eq-costlnmean-gradient})
should preferably 
be rewritten as%
\footnote{%
\ytextred{%
Each derivative
$
\frac{d 
\Yetwelveinfoinfominusconst
(
\ytextblue{\Yssoutsepsystsigdisctimecentvecrand}
)
}{d \Ysssepsystmatrixscalareloptim}
$
is "total" 
\ytextblue{only}
with respect to the considered coefficient
$
\Ysssepsystmatrixscalareloptim
$
(which is one of the two coefficients
\ysssepsystmatrixscalarelonetwo\
and
\ysssepsystmatrixscalareltwoone ),
i.e. it takes into account all variations of
$
\Yetwelveinfoinfominusconst
(
\Yssoutsepsystsigconttimecentvec
)
$
with respect to that coefficient
\ysssepsystmatrixscalareloptim\
while the other coefficient,
i.e.
$
{
\Ysssepsystmatrixscalarelnoindex _{\ell k}
}
$,
is kept constant.
For the sake of clarity, we could therefore
denote that derivative
$
\left(
\frac{d 
\Yetwelveinfoinfominusconst
(
\ytextblue{\Yssoutsepsystsigdisctimecentvecrand}
)
}{d \Ysssepsystmatrixscalareloptim}
\right)
_{
{
\Ysssepsystmatrixscalarelnoindex _{\ell k}
}
}
$,
\ytextblue{to show}
that
$
{
\Ysssepsystmatrixscalarelnoindex _{\ell k}
}
$ is constant.
However, this would decrease readability.
Therefore, in all this paper we omit 
\ytextblue{the notation
$
\left(
.
\right)
_{
{
\Ysssepsystmatrixscalarelnoindex _{\ell k}
}
}
$,}
but it should be kept
in mind that each considered derivative with respect to
\ysssepsystmatrixscalareloptim\ is calculated with
$
{
\Ysssepsystmatrixscalarelnoindex _{\ell k}
}
$
constant.
Then, in this framework, what we have to distinguish are:
(i) the total derivative due to the variations of
\ysssepsystmatrixscalareloptim ,
$
\ytextblue{\Yssoutsepsystsigdisctimecentonerand}
$
and
$
\ytextblue{\Yssoutsepsystsigdisctimecenttworand}
$
and
(ii) the partial derivative only due to
\ysssepsystmatrixscalareloptim .
We then have to use two different notations for these two types
of derivatives, such as
$
\frac{
d \Ysepfuncjacob
}{
d \Ysssepsystmatrixscalareloptim
}
$
and
$
\frac{
\partial \Ysepfuncjacob
}{
\partial \Ysssepsystmatrixscalareloptim
}
$
in
(\ref{eq-cost-deriv-total}).
This type of notations is commonly used in the literature
for functions which depend (i) on a single 
\ytextblue{independent variable},
i.e. time, and (ii) on other variables which themselves depend on
time, such as coordinate variables: see e.g.
http://en.wikipedia.org/wiki/Total\_derivative .
We here extend this concept to a configuration which involves
several 
\ytextblue{independent variables},
i.e.
\ysssepsystmatrixscalarelonetwo\
and
\ysssepsystmatrixscalareltwoone\
(and, again, other variables which themselves depend on
the 
\ytextblue{independent variables},
i.e.
$
\ytextblue{\Yssoutsepsystsigdisctimecentonerand}
$
and
$
\ytextblue{\Yssoutsepsystsigdisctimecenttworand}
$).
We keep the same type of notations as in the standard case involving
a single 
\ytextblue{independent variable}.
}
}
\begin{equation}
\label{eq-costlnmean-gradient-versiontwo}
\frac{d 
\Yetwelveinfoinfominusconst
(
\ytextblue{\Yssoutsepsystsigdisctimecentvecrand}
)
}{d \Ysssepsystmatrixscalareloptim}
=
\left(
\sum_{\Ysswayindex = 1}^{2}
E \{
\psi_{\Ysswayindex} (
\ytextblue{\Yssoutsepsystsigdisctimecentindexrand}
)
\frac{d
\ytextblue{\Yssoutsepsystsigdisctimecentindexrand}
}{d \Ysssepsystmatrixscalareloptim} \}
\right)
-
E \{
\frac{1}{\Ysepfuncjacob}
\frac{
d \Ysepfuncjacob
}{
d \Ysssepsystmatrixscalareloptim
}
\}
\end{equation}
still
with
(\ref{eq-score-func-def}).
The term
$
\frac{
d \Ysepfuncjacob
}{
d \Ysssepsystmatrixscalareloptim
}
$
in
(\ref{eq-costlnmean-gradient-versiontwo})
then deserves some care because, as shown by
(\ref{eq-mixfuncjacobval}), the Jacobian
\ysepfuncjacob\ contains 
the above-defined
two types of dependencies
with respect to 
\ysssepsystmatrixscalarelonetwo\
and
\ysssepsystmatrixscalareltwoone ,
i.e.
(i) \emph{direct dependencies} due to the factors in
(\ref{eq-mixfuncjacobval}) which explicitly contain
\ysssepsystmatrixscalarelonetwo\
and
\ysssepsystmatrixscalareltwoone\
and
(ii) \emph{indirect 
dependencies} due to the factors in
(\ref{eq-mixfuncjacobval}) which depend on 
$
\Yssoutsepsystsigdisctimecentone$
and
$
\Yssoutsepsystsigdisctimecenttwo
$,
which themselves depend on
\ysssepsystmatrixscalarelonetwo\
and
\ysssepsystmatrixscalareltwoone\
in this approach.
We here have to consider the \emph{total}
derivative
$
\frac{
d \Ysepfuncjacob
}{
d \Ysssepsystmatrixscalareloptim
}
$,
which takes into account both types of dependencies, and which
therefore reads
\begin{equation}
\label{eq-cost-deriv-total}
\frac{
d \Ysepfuncjacob
}{
d \Ysssepsystmatrixscalareloptim
}
=
\frac{
\partial \Ysepfuncjacob
}{
\partial \Ysssepsystmatrixscalareloptim
}
+
\sum_{\Ysswayindex = 1}^{2}
\frac
{\partial \Ysepfuncjacob}
{\partial \Yssoutsepsystsigdisctimecentindex}
\frac{d
\Yssoutsepsystsigdisctimecentindex
}{d \Ysssepsystmatrixscalareloptim}
.
\end{equation}
In this expression,
$
\displaystyle
\frac{
\partial \Ysepfuncjacob
}{
\partial \Ysssepsystmatrixscalareloptim
}
$
is the \emph{partial} derivative of
\ysepfuncjacob\ with respect to
\ysssepsystmatrixscalareloptim ,
calculated
by considering that the signals 
$
\Yssoutsepsystsigdisctimecentone$
and
$
\Yssoutsepsystsigdisctimecenttwo
$
are constant (in addition to the fact that the other
internal coefficient
\ysssepsystmatrixscalarelindexindexother\
of the separating system is also constant).
This 
partial derivative is 
the quantity
that is taken into account in 
the right-hand side of
(26) of
\cite{article-lduarte-mlsp2007}.
However, 
let us insist again that
this partial derivative
is first to be added with the other terms in the right-hand
side of
(\ref{eq-cost-deriv-total}),
in order to obtain
the overall
total
derivative
$
\displaystyle
\frac{
d \Ysepfuncjacob
}{
d \Ysssepsystmatrixscalareloptim
}
$
defined by
(\ref{eq-cost-deriv-total}).
What should eventually 
be used in the last term of
(\ref{eq-costlnmean-gradient}) or
(\ref{eq-costlnmean-gradient-versiontwo})
is this \emph{total} derivative.

So, starting from the expression of
\ysepfuncjacob\ provided in
(\ref{eq-mixfuncjacobval}), one easily derives all
its partial derivatives involved in
(\ref{eq-cost-deriv-total}).
They read as follows
\begin{eqnarray}
\frac{
\partial \Ysepfuncjacob
}{
\partial \Ysssepsystmatrixscalarelonetwo
}
&
=
&
\frac{
\Ysssepsystmatrixscalareltwoone
\Yssoutsepsystsigdisctimecentone
^{
\frac{1}{\Ymixfuncexponent} - 1
}
\Yssoutsepsystsigdisctimecenttwo
^{
\Ymixfuncexponent -1
}
}
{
[
1
-
\Ysssepsystmatrixscalarelonetwo
\Ysssepsystmatrixscalareltwoone
\Yssoutsepsystsigdisctimecentone
^{
\frac{1}{\Ymixfuncexponent} - 1
}
\Yssoutsepsystsigdisctimecenttwo
^{
\Ymixfuncexponent -1
}
]^2
}
\\
\frac{
\partial \Ysepfuncjacob
}{
\partial \Ysssepsystmatrixscalareltwoone
}
&
=
&
\frac{
\Ysssepsystmatrixscalarelonetwo
\Yssoutsepsystsigdisctimecentone
^{
\frac{1}{\Ymixfuncexponent} - 1
}
\Yssoutsepsystsigdisctimecenttwo
^{
\Ymixfuncexponent -1
}
}
{
[
1
-
\Ysssepsystmatrixscalarelonetwo
\Ysssepsystmatrixscalareltwoone
\Yssoutsepsystsigdisctimecentone
^{
\frac{1}{\Ymixfuncexponent} - 1
}
\Yssoutsepsystsigdisctimecenttwo
^{
\Ymixfuncexponent -1
}
]^2
}
\\
\label{eq-deriv-sepfuncjacob-vs-outsepsystsigdisctimecentone}
\frac{
\partial \Ysepfuncjacob
}{
\partial 
\Yssoutsepsystsigdisctimecentone
}
&
=
&
\frac{
\Ysssepsystmatrixscalarelonetwo
\Ysssepsystmatrixscalareltwoone
\left(
\frac{1}{\Ymixfuncexponent} - 1
\right)
\Yssoutsepsystsigdisctimecentone
^{
\frac{1}{\Ymixfuncexponent} - 2
}
\Yssoutsepsystsigdisctimecenttwo
^{
\Ymixfuncexponent -1
}
}
{
[
1
-
\Ysssepsystmatrixscalarelonetwo
\Ysssepsystmatrixscalareltwoone
\Yssoutsepsystsigdisctimecentone
^{
\frac{1}{\Ymixfuncexponent} - 1
}
\Yssoutsepsystsigdisctimecenttwo
^{
\Ymixfuncexponent -1
}
]^2
}
\\
\label{eq-deriv-sepfuncjacob-vs-outsepsystsigdisctimecenttwo}
\frac{
\partial \Ysepfuncjacob
}{
\partial 
\Yssoutsepsystsigdisctimecenttwo
}
&
=
&
\frac{
\Ysssepsystmatrixscalarelonetwo
\Ysssepsystmatrixscalareltwoone
\Yssoutsepsystsigdisctimecentone
^{
\frac{1}{\Ymixfuncexponent} - 1
}
\left(
\Ymixfuncexponent - 1
\right)
\Yssoutsepsystsigdisctimecenttwo
^{
\Ymixfuncexponent -2
}
}
{
[
1
-
\Ysssepsystmatrixscalarelonetwo
\Ysssepsystmatrixscalareltwoone
\Yssoutsepsystsigdisctimecentone
^{
\frac{1}{\Ymixfuncexponent} - 1
}
\Yssoutsepsystsigdisctimecenttwo
^{
\Ymixfuncexponent -1
}
]^2
}
.
\end{eqnarray}
The case when
$
\Ymixfuncexponent
=
1
$
deserves a comment.
As shown by
(\ref{eq-mixsigconttimecentoneval})-%
(\ref{eq-mixsigconttimecenttwoval}),
the mixing model then becomes
linear.
Besides,
as shown by
(\ref{eq-deriv-sepfuncjacob-vs-outsepsystsigdisctimecentone})-%
(\ref{eq-deriv-sepfuncjacob-vs-outsepsystsigdisctimecenttwo}),
we then
have
\begin{eqnarray}
\frac{
\partial \Ysepfuncjacob
}{
\partial 
\Yssoutsepsystsigdisctimecentone
}
&
=
&
0
\\
\frac{
\partial \Ysepfuncjacob
}{
\partial 
\Yssoutsepsystsigdisctimecenttwo
}
&
=
&
0 ,
\end{eqnarray}
so that the total derivative
$
\frac{
d \Ysepfuncjacob
}{
d \Ysssepsystmatrixscalareloptim
}
$
in
(\ref{eq-cost-deriv-total})
becomes equal to the partial
derivative
$
\frac{
\partial \Ysepfuncjacob
}{
\partial \Ysssepsystmatrixscalareloptim
}
$
in
(\ref{eq-cost-deriv-total}).
This clearly shows that the problems due to the
distinction between these two derivatives, that we address
in this paper, concern \emph{nonlinear} mixtures.

The last terms which are required to obtain the
complete expressions in
(\ref{eq-costlnmean-gradient-versiontwo})%
\footnote{Eq.
(\ref{eq-costlnmean-gradient-versiontwo})
is obtained by
taking the derivative of
(\ref{eq-ln-pdf-mix-vs-src-versiontwo})
with respect to
\ysssepsystmatrixscalareloptim .
It thus relies on the fact that
$
\frac{d 
\YetwelveinfoRVcontentropydiff
(
\ytextblue{\Yssoutsepsystsigdisctimecentindexrand}
)
}{d \Ysssepsystmatrixscalareloptim}
=
E \{
\psi_{\Ysswayindex} (
\ytextblue{\Yssoutsepsystsigdisctimecentindexrand}
)
\frac{d
\ytextblue{\Yssoutsepsystsigdisctimecentindexrand}
}{d \Ysssepsystmatrixscalareloptim} \}
$.
In \cite{article-lduarte-mlsp2007},
this result was borrowed from 
\cite{article-taleb-jutten}.
Considering the problems due to indirect dependencies
in nonlinear mixtures
found in
\cite{article-lduarte-mlsp2007},
one may wonder whether the relationship
$
\frac{d 
\YetwelveinfoRVcontentropydiff
(
\ytextblue{\Yssoutsepsystsigdisctimecentindexrand}
)
}{d \Ysssepsystmatrixscalareloptim}
=
E \{
\psi_{\Ysswayindex} (
\ytextblue{\Yssoutsepsystsigdisctimecentindexrand}
)
\frac{d
\ytextblue{\Yssoutsepsystsigdisctimecentindexrand}
}{d \Ysssepsystmatrixscalareloptim} \}
$
still holds for the nonlinear mixing model studied in
\cite{article-lduarte-mlsp2007}.
We claim that it does hold.%
}
and
(\ref{eq-cost-deriv-total})
are \emph{all four derivatives}
$
\frac{d
\Yssoutsepsystsigdisctimecentindex
}{d \Ysssepsystmatrixscalareloptim}
$.
For the sake of clarity, we now show how they may
be considered, when taking into account the above
comments about total and partial derivatives.
Here again, 
\ysssepsystmatrixscalarelonetwo\
and
\ysssepsystmatrixscalareltwoone\
should be considered as the independent variables,
while
\yssoutsepsystsigdisctimecentone\
and
\yssoutsepsystsigdisctimecenttwo\
are functions of them and the observations
are constant.
All these parameters are linked by
(\ref{eq-outsepsystsigdisctimecentone-converg})-%
(\ref{eq-outsepsystsigdisctimecenttwo-converg}).
By first
computing the total derivatives of the latter equations
with respect to
\ysssepsystmatrixscalarelonetwo ,
one gets
\begin{eqnarray}
\label{eq-outsepsystsigdisctimecentone-converg-total-deriv}
\frac{d
\Yssoutsepsystsigdisctimecentone
}
{d \Ysssepsystmatrixscalarelonetwo}
&
=
&
-
(
\Yssoutsepsystsigdisctimecenttwo
^{
\Ymixfuncexponent
}
+
\Ysssepsystmatrixscalarelonetwo
\Ymixfuncexponent
\Yssoutsepsystsigdisctimecenttwo
^{
\Ymixfuncexponent -1
}
\frac{d
\Yssoutsepsystsigdisctimecenttwo
}
{d \Ysssepsystmatrixscalarelonetwo}
)
\\
\label{eq-outsepsystsigdisctimecenttwo-converg-total-deriv}
\frac{d
\Yssoutsepsystsigdisctimecenttwo
}
{d \Ysssepsystmatrixscalarelonetwo}
&
=
&
-
\Ysssepsystmatrixscalareltwoone
\frac{1}{\Ymixfuncexponent}
\Yssoutsepsystsigdisctimecentone
^{
\frac{1}{\Ymixfuncexponent} - 1
}
\frac{d
\Yssoutsepsystsigdisctimecentone
}
{d \Ysssepsystmatrixscalarelonetwo}
.
\end{eqnarray}
Inserting
(\ref{eq-outsepsystsigdisctimecenttwo-converg-total-deriv})
in
(\ref{eq-outsepsystsigdisctimecentone-converg-total-deriv}),
one derives
\begin{equation}
\label{eq-outsepsystsigdisctimecentone-converg-total-deriv-extract}
\frac{d
\Yssoutsepsystsigdisctimecentone
}
{d \Ysssepsystmatrixscalarelonetwo}
=
\frac{
-
\Yssoutsepsystsigdisctimecenttwo
^{
\Ymixfuncexponent
}
}
{
1
-
\Ysssepsystmatrixscalarelonetwo
\Ysssepsystmatrixscalareltwoone
\Yssoutsepsystsigdisctimecentone
^{
\frac{1}{\Ymixfuncexponent} - 1
}
\Yssoutsepsystsigdisctimecenttwo
^{
\Ymixfuncexponent -1
}
}
.
\end{equation}
Then inserting
(\ref{eq-outsepsystsigdisctimecentone-converg-total-deriv-extract})
in
(\ref{eq-outsepsystsigdisctimecenttwo-converg-total-deriv}),
one obtains
\begin{equation}
\frac{d
\Yssoutsepsystsigdisctimecenttwo
}
{d \Ysssepsystmatrixscalarelonetwo}
=
\frac
{
\Ysssepsystmatrixscalareltwoone
\frac{1}{\Ymixfuncexponent}
\Yssoutsepsystsigdisctimecentone
^{
\frac{1}{\Ymixfuncexponent} - 1
}
\Yssoutsepsystsigdisctimecenttwo
^{
\Ymixfuncexponent
}
}
{
1
-
\Ysssepsystmatrixscalarelonetwo
\Ysssepsystmatrixscalareltwoone
\Yssoutsepsystsigdisctimecentone
^{
\frac{1}{\Ymixfuncexponent} - 1
}
\Yssoutsepsystsigdisctimecenttwo
^{
\Ymixfuncexponent -1
}
}
.
\end{equation}
Similarly,
computing the total derivatives of 
(\ref{eq-outsepsystsigdisctimecentone-converg})-%
(\ref{eq-outsepsystsigdisctimecenttwo-converg})
with respect to
\ysssepsystmatrixscalareltwoone\
eventually 
yields
\begin{eqnarray}
\frac{d
\Yssoutsepsystsigdisctimecentone
}
{d \Ysssepsystmatrixscalareltwoone}
&
=
&
\frac
{
\Ysssepsystmatrixscalarelonetwo
{\Ymixfuncexponent}
\Yssoutsepsystsigdisctimecentone
^{
\frac{1}{\Ymixfuncexponent}
}
\Yssoutsepsystsigdisctimecenttwo
^{
\Ymixfuncexponent -1
}
}
{
1
-
\Ysssepsystmatrixscalarelonetwo
\Ysssepsystmatrixscalareltwoone
\Yssoutsepsystsigdisctimecentone
^{
\frac{1}{\Ymixfuncexponent} - 1
}
\Yssoutsepsystsigdisctimecenttwo
^{
\Ymixfuncexponent -1
}
}
\\
\frac{d
\Yssoutsepsystsigdisctimecenttwo
}
{d 
\ytextred
{\Ysssepsystmatrixscalareltwoone}
}
&
=
&
\frac{
-
\Yssoutsepsystsigdisctimecentone
^{
\frac{1}{\Ymixfuncexponent}
}
}
{
1
-
\Ysssepsystmatrixscalarelonetwo
\Ysssepsystmatrixscalareltwoone
\Yssoutsepsystsigdisctimecentone
^{
\frac{1}{\Ymixfuncexponent} - 1
}
\Yssoutsepsystsigdisctimecenttwo
^{
\Ymixfuncexponent -1
}
}
.
\end{eqnarray}
The expressions of
all four derivatives
$
\frac{d
\Yssoutsepsystsigdisctimecentindex
}{d \Ysssepsystmatrixscalareloptim}
$
obtained with this approach remain equal to the expressions
(30)-(33) of
\cite{article-lduarte-mlsp2007}, except that all 
partial derivative
\emph{notations}
$
\frac{\partial
\Yssoutsepsystsigdisctimecentindex
}{\partial \Ysssepsystmatrixscalareloptim}
$
in
\cite{article-lduarte-mlsp2007}
are here 
replaced by total derivative notations
$
\frac{d
\Yssoutsepsystsigdisctimecentindex
}{d \Ysssepsystmatrixscalareloptim}
$.

Gathering all above expressions
then makes it possible to determine the total
derivative
$
\frac{
d \Ysepfuncjacob
}{
d \Ysssepsystmatrixscalareloptim
}
$
in
(\ref{eq-cost-deriv-total}),
and then the overall gradient components in
(\ref{eq-costlnmean-gradient-versiontwo}).
This yields
the correct expression of the
gradient of the considered 
BSS criterion
based on
mutual information.

This correct gradient expression
may eventually be used to optimize
the adaptive coefficients
\ysssepsystmatrixscalarelonetwo\
and
\ysssepsystmatrixscalareltwoone ,
e.g. using the well-known gradient descent
algorithm.


\begin{thebibliography}{99}
\bibitem{amoi5-32}
Y. Deville, S. Hosseini,
"Stable Higher-Order Recurrent Neural Network Structures for Nonlinear 
Blind Source Separation",
Proceedings of the 7th
International Conference on
Independent Component Analysis and 
Signal
Separation
(ICA 2007),
pp. 161-168,
ISSN 0302-9743,
Springer-Verlag,
vol. LNCS 4666,
London, UK, September 9-12, 2007.
\bibitem{amoi5-49}
Y. Deville, S. Hosseini,
"Recurrent networks for separating extractable-target 
nonlinear mixtures. Part I: non-blind configurations",
Signal Processing,
vol. 89, no. 4, pp. 378-393, April 2009.
http://dx.doi.org/10.1016/j.sigpro.2008.09.016
\bibitem{bibref-ellipses-chap-mv-moi-arxiv}
Y. Deville, A. Deville,
"Effect of indirect dependencies on
"Maximum likelihood blind separation of two quantum states (qubits)
with cylindrical-symmetry Heisenberg spin coupling"",
http://arxiv.org/abs/0906.0062
\bibitem{article-lduarte-mlsp2007}
L. T. Duarte, C. Jutten. "A mutual information minimization approach for a class of nonlinear recurrent separating systems",
IEEE International Workshop on Machine Learning for Signal Processing, Thessaloniki, Greece, 2007. 
\bibitem{amoi4-39}
S. Hosseini, Y. Deville,
"Blind separation of linear-quadratic mixtures of real sources using a
recurrent structure",
Proceedings of the 7th International Work-conference on Artificial
And Natural Neural Networks (IWANN 2003),
special session,
vol. 2,
pp. 241-248,
J. Mira and J. R. Alvarez eds (Springer),
Mao, Menorca, Spain, June 3-6, 2003.
\bibitem{amoi4-48}
S. Hosseini, Y. Deville,
"Blind maximum likelihood separation of a linear-quadratic
mixture",
Proceedings of the Fifth International Conference on
Independent Component Analysis and Blind Signal Separation
(ICA 2004),
pp. 694-701,
ISSN 0302-9743,
ISBN 3-540-23056-4,
Springer-Verlag,
vol. LNCS
3195,
Granada, Spain, Sept. 22-24, 2004.
Springer-Verlag on-line version:
http://www.springerlink.com/index/J91PEDUGYCMDQGHD
\bibitem{icabook-oja}
A. Hyv\"arinen, J. Karhunen, E. Oja,
"Independent Component Analysis",
Wiley, New York, 2001.
\bibitem{article-taleb-jutten}
A. Taleb, C. Jutten,
"Source separation in post-nonlinear mixtures",
IEEE Transactions on signal processing,
vol. 47, no. 10, pp. 2807-2820,
Oct. 1999.
\end{thebibliography}
\end{document}